\documentclass[a4paper,11pt]{article}
\usepackage{jheppub} 
\usepackage{amsmath, amssymb, amsthm, epsfig, fancyhdr,epsfig,multirow}
\usepackage[utf8]{inputenc}
\usepackage{amsmath}
\usepackage{amsfonts}
\usepackage{amssymb}
\usepackage{xcolor}
\usepackage{comment}
\usepackage{subfig}
\usepackage[normalem]{ulem}
\usepackage{tabularx}
\usepackage{comment}
\usepackage{float}
\usepackage{graphicx}
\usepackage{appendix}
\usepackage[most]{tcolorbox}
\usepackage{physics}

\title{Bispectrum at 1-loop in the Effective Field Theory of Inflation}

\author{Supritha Bhowmick, Diptimoy Ghosh, Farman Ullah}
\affiliation{Indian Institute of Science Education and Research Pune, Pune 411008, India}

\emailAdd{supritha.bhowmick@students.iiserpune.ac.in} 
\emailAdd{diptimoy.ghosh@gmail.com} 
\emailAdd{farman.ullah@students.iiserpune.ac.in}

\abstract{In this paper we compute 1-loop corrections to the bispectrum in the decoupling limit of the Effective Field Theory of Inflation (EFToI). We regulate the divergences by employing dimensional regularization and work in $d=3+\delta$ dimensions. We find that the final results feature analytic structures of the form $\log{\left(k_i/k_T\right)}$ and $\log{\left(H/\mu\right)}$, where $H$ is the Hubble parameter and $\mu$ is the renormalisation scale. An interesting outcome of our calculations is that unlike the 1-loop correction to the power-spectrum computed in \cite{Senatore:2009cf} the unrenormalised answers always produce unphysical logarithms of co-moving momenta. These unphysical logarithms are cancelled only after renormalisation. We expect this to be a generic feature for loop computations unless there is some cancellation as in the previously computed 1-loop result for the power-spectrum.}

\begin{document}
\maketitle
\flushbottom

\section{Introduction}
The current leading candidate among the theories of the early universe is an epoch of inflation \cite{Starobinsky:1980te,Guth:1980zm,Linde:1981mu} when the universe was driven to spatial flatness, homogeneity and isotropy on large scales. It is believed that small quantum fluctuations in the inflaton field are the seeds of cosmological structures and inhomogeneities in the CMB \cite{Starobinsky:1979ty,Mukhanov:1981xt}. There has been a lot of progress in computing various correlation functions, mostly at the tree level, during inflation, since they can be mapped to cosmological observations, such as correlations in the CMB temperature fluctuations. A lot of tree-level formalism has been developed to understand the structure of these correlators in gory detail\cite{Maldacena:2002vr,Arkani-Hamed:2018kmz,Baumann:2019oyu,Baumann:2020dch,Pajer:2020wnj,Pajer:2020wxk,Jazayeri:2021fvk,DuasoPueyo:2020ywa,Ghosh:2022cny,Ghosh:2023agt}.\par
Loop corrections to cosmological correlation functions were first studied by Weinberg \cite{Weinberg:2005vy}, where he computed the 1-loop correction to the 2-point correlation function. However, it was pointed out in \cite{Senatore:2009cf} that there were mistakes in the computation, since the loop correction computed in \cite{Weinberg:2005vy} and subsequent related studies \cite{Seery:2007we,Chaicherdsakul:2006ui,Adshead:2009cb,Gao:2009fx,Campo:2009fx} have a $\text{log}(k/ \mu)$ dependence. Such a term is unphysical because it is the ratio of the co-moving momentum $k$ and the physical renormalization scale $\mu$. In \cite{Senatore:2009cf} it was shown that once the mode functions are properly continued to $d$ dimensions in dimensional regularization, the unphysical logarithms cancel and the log dependence instead is $\text{log}(H/ \mu)$. Loop effects were studied in further detail in \cite{Senatore:2012nq,Pimentel:2012tw}.

It is crucial to go beyond the 1-loop 2-point function to capture features of interactions in the shape of the correlation function. Moreover, the analytic structure of higher point cosmological correlators at the loop level is mostly unexplored. It therefore demands a thorough theoretical investigation. In this paper, we study 1-loop corrections to the scalar 3-point function in the decoupling limit of the Effective Field Theory of Inflation (EFToI). Since our main objective is to explore the analytic structure of the correlation functions, we perform our calculations for a tuned scenario (see Sec. \ref{3}) reducing the number of possible Feynmann diagrams. We find a few interesting features in our final results:\\
\begin{itemize}
\item The correlation function is not a rational function unlike the parity odd correlators computed in \cite{
Lee:2023jby}.

\item Unlike \cite{Senatore:2009cf}, the unphysical logs do not cancel despite using correct $d$ dimensional mode functions. They are removed only after renormalizing the results. 

\item Our final renormalized result for the correlation function contains logarithmic structures of the form $\log{\frac{k_i}{k_T}}$ and $\log{\frac{H}{\mu}}$, where $k_T=k_1+k_2+k_3$.
\end{itemize}

The rest of the paper is organised as follows. In Sec. \ref{3} we briefly revisit the decoupling limit of EFToI and explicitly write down the tuned action we use for computations. In Sec. \ref{4} we give details of the loop computations. In Sec. \ref{5} we renormalise our answers and show that unphysical logarithms cancel. In Sec. \ref{Sec.cutoff} we perform a cross check where we compute one of our diagrams in cutoff regularisation. Finally, we conclude in Sec. \ref{6}.

\section{Notations and Conventions}
We will compute loop corrections to correlations of fluctuations on the de-Sitter background, given by
\begin{align}
    ds^2 =\frac{ -d\tau^2 + d\vec{x}^2}{\left(H\tau\right)^{2}}\,,
\end{align}
 where $\tau$ is the conformal time and $H$ is the Hubble parameter. This is a good approximation to the quasi-de Sitter inflationary background.
We will work with the interaction Hamiltonian $H_{I}$ obtained from the EFToI in Eq.~\eqref{eqn:EFToI}. In the computations that follow, we denote internal (loop) momenta by $p_i$ and external momenta by $k_i$. Derivatives with respect to physical time are denoted by dots ($\cdot$), and the ones with respect to conformal time are primed ($'$). The divergences arising in our loop integrals are regulated by dimensional regularization. The mode functions in $d$ dimensions for massive fields are Hankel functions, $H^{(1)}_{\nu}(-k\tau)$, where $\nu=\sqrt{\frac{d^{2}}{4}-\frac{m^{2}}{H^{2}}}$. Even for the massless case, the mode functions are complicated Hankel functions $H^{(1)}_{d/2}(-k\tau)$ in $d$ dimensions (unlike $d=3$, where the mode functions are simpler to work with), which makes the loop computations difficult. Instead, we implement a slightly modified form of dimensional regularization by using the trick \cite{Melville:2021lst,Lee:2023jby} of analytically continuing the mass to $d$ dimensions, i.e. $m^2 \rightarrow \frac{H^{2}}{4}\left(d^{2}-9\right)$. In this case, we get $\nu=\frac{3}{2}$ and $m\rightarrow 0$ when $d\rightarrow 3$. This gives back simple massless mode functions of the form,
\begin{align}
      f_{k}(\tau)=\left(-H\tau\right)^{\delta/2}\frac{H}{\sqrt{-\dot{H}M_{pl}^2}} \frac{1}{\sqrt{2k^{3}}}\left(1-ik\tau\right)e^{ik\tau}\,.
\label{mode_funcs}
\end{align}

\section{Decoupling limit of EFToI} \label{3}
The EFToI~\cite{Cheung:2007st,Weinberg:2008hq,Piazza:2013coa} lagrangian is constructed in the unitary gauge by fixing time diffeomorphisms, and therefore no longer invariant under time diffs. One can restore this invariance by introducing an additional field $\pi(x)$ in the lagrangian using the so-called Stueckelberg trick. The action for $\pi(x)$ is given by,
\begin{align}
    S=\int d^{4}x\sqrt{-g}& \big[\frac{1}{2}M^{2}_{pl}R-M^{2}_{pl}\left(3H^{2}(t+\pi)+\dot{H}(t+\pi)\right)+ \nonumber \\
    & M^{2}_{pl}\dot{H}(t+\pi)\left((1+\dot{\pi})^{2}g^{00}+2(1+\dot{\pi})\partial_{i}\pi g^{0i}+g^{ij}\partial_i\pi\partial_j\pi+1\right)+ \nonumber \\
       & \frac{M_2(t+\pi)^{4}}{2!} \left((1+\dot{\pi})^{2}g^{00}+2(1+\dot{\pi})\partial_{i}\pi g^{0i}+g^{ij}\partial_i\pi\partial_j\pi+1\right)^2+ \nonumber \\
        & \frac{M_3(t+\pi)^{4}}{3!}\left((1+\dot{\pi})^{2}g^{00}+2(1+\dot{\pi})\partial_{i}\pi g^{0i}+g^{ij}\partial_i\pi\partial_j\pi+1\right)^3+ \nonumber \\
        & \frac{M_4(t+\pi)^{4}}{4!}\left((1+\dot{\pi})^{2}g^{00}+2(1+\dot{\pi})\partial_{i}\pi g^{0i}+g^{ij}\partial_i\pi\partial_j\pi+1\right)^4 + .....\big]\,. 
\end{align}
This action is invariant under the transformation, $t \rightarrow t + \epsilon(x)$ \& $\pi\rightarrow \pi(x) - \epsilon(x)$. One can define a limit of the above action where the energy scale associated with the coupling between metric and scalar fluctuations, $E_\text{mix}$ is much smaller than the energy scale of the processes under consideration, i.e. $E \gg E_\text{mix}$. This is known as the decoupling limit of EFToI. If the observables we wish to compute are not dominated by mixing with gravity, the action in the decoupling limit will give the correct result. This is very useful because the above action in the decoupling limit for the scalar sector simplifies to,
 \begin{align} \label{EFToI}
    S=\int d^{4}x\sqrt{-g}& \big[\frac{1}{2}M^{2}_{pl}R-M^{2}_{pl}\left(3H^{2}(t+\pi)+\dot{H}(t+\pi)\right)+ \nonumber \\
     & M^{2}_{pl}\dot{H}(t+\pi)\left(-(1+\dot{\pi})^{2}+\frac{1}{a^{2}}(\partial_i \pi)^{2}+1\right)+ \nonumber \\
       & \frac{1}{2!}M_2(t+\pi)^{4}\left(-(1+\dot{\pi})^{2}+\frac{1}{a^{2}}(\partial_i \pi)^{2}+1\right)^{2}+ \nonumber \\
        &\frac{1}{3!} M_3(t+\pi)^{4}\left(-(1+\dot{\pi})^{2}+\frac{1}{a^{2}}(\partial_i \pi)^{2}+1\right)^{3}+ \nonumber \\
       & \frac{1}{4!} M_4(t+\pi)^{4}\left(-(1+\dot{\pi})^{2}+\frac{1}{a^{2}}(\partial_i \pi)^{2}+1\right)^{4}+ .....\big]\,. 
\end{align}
We assume that the time dependence of EFT coefficients is slow-roll suppressed and therefore can be neglected in the first approximation. Since the goal of this paper is to explore the general analytic structure for the loop-level bispectrum, we consider a scenario where the coefficients of the EFT are tuned, such that only one cubic and two quartic interactions are important. Then we have the following action for $\pi$,
\begin{align}
    S= \int d^{4}x \sqrt{-g}\bigg[-M^{2}_{pl}\dot{H}\left(\dot{\pi}^{2}-\frac{(\partial\pi)^{2}}{a^{2}}\right)+g_3\dot{\pi}^{3} + g_{4}\dot{\pi}^{4}-\frac{3 g_3}{2}\dot{\pi}^{2}\left(\frac{\partial\pi}{a}\right)^{2}\bigg]\,,
    \label{eqn:EFToI}
\end{align}
where we have set $M_2=0$ and 
\begin{align}
  &g_3=-\frac{4}{3}M^{4}_3\,, \\
  &g_{4}=-2M^{4}_3+\frac{2}{3}M^{4}_4\,.\\ \nonumber
\end{align}
\section{Explicit computation of Bispectrum at 1-loop}\label{4}
In this section, we compute the 1-loop contribution to the 3-point correlator coming from the leading cubic and quartic operators in the EFToI, Eq.~\eqref{eqn:EFToI}. The computations are done using the in-in formalism (reviewed in Appendix \ref{in:in}), 
\begin{eqnarray}
    \langle \pi^3 (0) \rangle = \frac{\langle 0 | \bar{T} e^{i\int^0_{-\infty(1+i \epsilon)} H_I d\tau}~ {\pi}^{3} (0)~ T e^{-i \int^0_{-\infty(1-i\epsilon)}H_I d\tau} |0\rangle}{\langle 0 | \bar{T} e^{i\int^0_{-\infty(1+i \epsilon)} H_I d\tau}~ T e^{-i \int^0_{-\infty(1-i\epsilon)}H_I d\tau} |0\rangle}\,,
    \label{Eqn:in-in}
\end{eqnarray}
where the correlation functions are computed at the future boundary of de Sitter, i.e. $\tau \rightarrow 0$. Here ($\Bar{T}$) $T$ denotes (anti-)time ordering. The time integrals are performed using the usual continuation $\tau\rightarrow \tau(1 \pm i \epsilon)$, which rotates the time contour and projects the free vacuum on the interacting vacuum. 

\begin{figure}
\centering
\subfloat[][]{\includegraphics[scale=0.55]{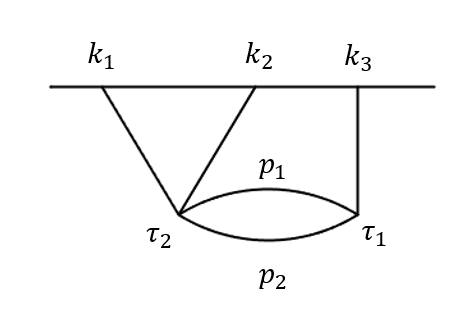}\label{Fig:O(c_3 c_4)_1}}
\subfloat[][]{\includegraphics[scale=0.55]{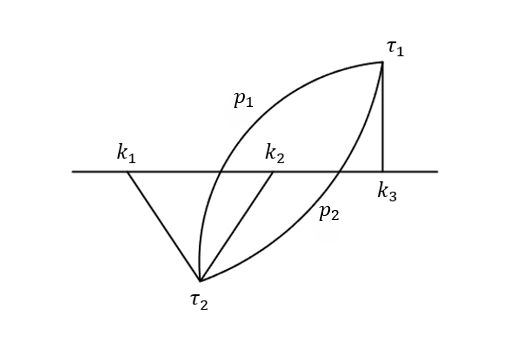}\label{Fig:O(c_3 c_4)_2}}
\caption{Diagram I : Loop contribution to the bispectrum from one insertion of the cubic and a quartic operator. $k_1$ , $k_2$ and $k_3$ denote external momenta. $p_1$ and $p_2$ denote loop momenta. $\tau_1$ and $\tau_2$ are points in the bulk. Time ordered diagram is shown in the left panel and the diagram with no specific time ordering is drawn in the right panel.}
\label{Fig:O(c_3 c_4)}
\end{figure}

\begin{figure}
\centering
\subfloat[][]{\includegraphics[scale=0.55]{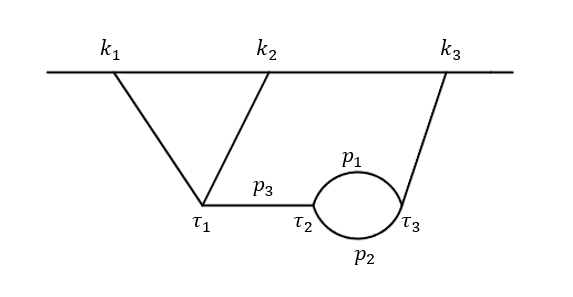}\label{Fig:O(c_3^3)_1}}
\subfloat[][]{\includegraphics[scale=0.55]{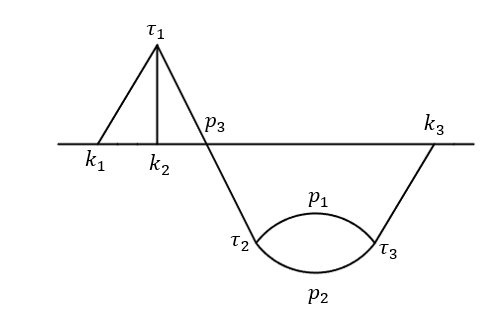}\label{Fig:O(c_3^3)_2}} \\
\subfloat[][]{\includegraphics[scale=0.55]{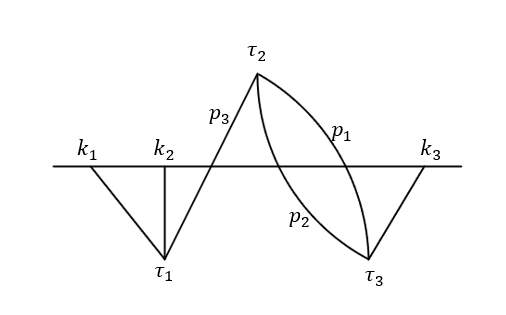}\label{Fig:O(c_3^3)_3}}
\subfloat[][]{\includegraphics[scale=0.55]{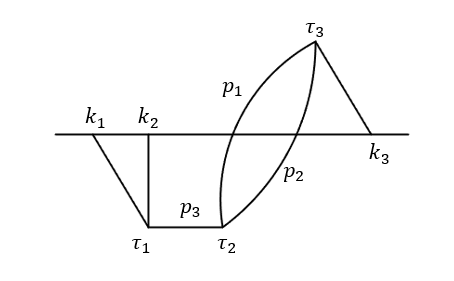}\label{Fig:O(c_3^3)_4}}
\caption{Diagram II : Loop contribution to the bispectrum from three insertions of the cubic operator. $k_1$ , $k_2$ and $k_3$ denote external momenta. $p_1$ , $p_2$ are loop momenta and $p_3=k_3$ due to momentum conservation. $\tau_1$ , $\tau_2$ and $\tau_3$ are points in the bulk.}
\label{Fig:O(c_3^3)}
\end{figure}

\begin{figure}
\centering
\subfloat[][]{\includegraphics[scale=0.7]{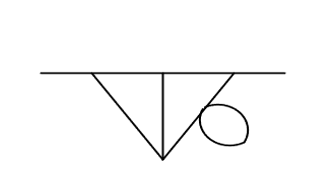}\label{Fig:scaleless_1}}
\subfloat[][]{\includegraphics[scale=0.55]{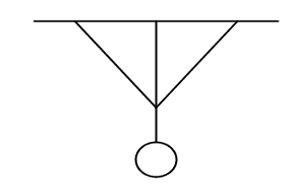}\label{Fig:scaleless_2}} \\
\caption{Diagrams with scaleless loops (left panel) and tadpoles (right panel). Contributions from scaleless loops vanish in dimensional regularization. Since all the interactions have derivatives, the tadpole diagram integrand either vanishes, by virtue of the tadpole leg $k=0$, or is finite and its contribution vanishes from scaleless loop arguments (see Sec.~\ref{Scaleless_integrals})}
\label{Fig:scalelessloop}
\end{figure}

The 1-loop corrections can arise from one insertion of a cubic and a quartic operator each, given by Fig.~\ref{Fig:O(c_3 c_4)}, which we refer to as Diagram I. In Sec.~\ref{Dia_1_computation}, we provide details of their computation. All other contributions from a cubic and a quartic interaction are different time orderings of Fig.~\ref{Fig:O(c_3 c_4)}, obtained by complex conjugation of the same. At the same order in the EFT expansion, one can also have contributions from diagrams with three insertions of the cubic operator. Some of these diagrams are given in Fig.~\ref{Fig:O(c_3^3)}, which we refer to as Diagram II. We calculate Fig.~\ref{Fig:O(c_3^3)_1} in Sec.~\ref{Dia_2_computation} and show that the analytic structure is similar to Diagram I, leaving an explicit computation of all other diagrams for the future. 
Finally, there are contributions from diagrams where the loop closes on itself, including tadpole diagrams, resulting in a scaleless integral. Such diagrams are given in Fig.~\ref{Fig:scalelessloop}. These contributions vanish in dim-reg, as discussed in Sec. \ref{Scaleless_integrals}.

\subsection{Calculational scheme} \label{section:calc_Scheme}
Since we will compute many diagrams using the same method, it is useful to first give the general calculational outline and then proceed to compute explicitly. To compute any diagram $D(k_1,k_2,k_3)$, we use the well-known Feynman rules \cite{Chen:2017ryl}. When working in $d=3+\delta$ dimensions, the mode function (Eq. \eqref{mode_funcs}) and the measure will have $\delta$ dependence, given by
\begin{align}
    & f_{k}(\tau)=\left(-H\tau\right)^{\delta/2}\frac{H}{\sqrt{-\dot{H}M_{pl}^2}} \frac{1}{\sqrt{2k^{3}}}\left(1-ik\tau\right)e^{ik\tau} = \left(1+\frac{\delta}{2} \log \left(-H \tau \right) \right) f_k^{3d}(\tau) + \mathcal{O}(\delta^2) \,, \nonumber \\
    &  a(\tau)^{\delta} = 1 - \delta \text{log} (-H \tau) + \mathcal{O}(\delta^2) \,.
    \label{eqn:delta_contr}
\end{align}

In our computations, the external mode functions (at $\tau =0$) and the external momenta conserving delta function ($\delta^d (\sum_{i} \vec{k}_i)$) are taken to be in 3 dimensions, since the counter terms cancel the $O(\delta)$ contribution from these terms. Our integrals have the following form,
\begin{align}
   & D(k_1,k_2,k_3)=\delta^3\big(\sum_i \vec{k}_i \big)~\mu^{-\frac{3 \delta}{2}}  \int d^{d}\vec{p}_1 d^{d}\vec{p}_2~P(p_1,p_2,\{k_i\},\{\tau_i\})\delta^{(d)}(\vec{p}_1+\vec{p}_2-\vec{k}_3)\nonumber \\
    &  \hspace{200pt} \times \prod_i \left[(-H\tau_i)^{(n_i)\delta}d\tau_i \right]\,,\nonumber \\
   &   D(k_1,k_2,k_3)=\delta^3\big(\sum_i \vec{k}_i \big)~ \mu^{-\frac{3 \delta}{2}} \int d^{d}\vec{p}_1 d^{d}\vec{p}_2P(p_1,p_2,\{k_i\},\{\tau_i\})\delta^{(d)}(\vec{p}_1+\vec{p}_2-\vec{k}_3) \nonumber \\
    &  \hspace{200pt} \times \left[1+\sum_{i}n_i\delta\log(-H\tau_i)\right]\prod_{i}d\tau_i \,,
    \end{align}
   where $\mu$ is the renormalisation scale.
   In passing we note that From Eq. \eqref{eqn:delta_contr} we get $n_i=-1$ from each scale factor, $a(\tau_i)^{\delta}$, and $n_i=\frac{1}{2}$ from each  mode function. Now, we can write $D=D_1+D_2$, with,
    \begin{align}
  &  D_1(k_1,k_2,k_3)=\delta^3\big(\sum_i \vec{k}_i \big)~\mu^{-\frac{3 \delta}{2}}  \int d^{d}\vec{p}_1 d^{d}\vec{p}_2P(p_1,p_2,\{k_i\},\{\tau_i\})\delta^{(d)}(\vec{p}_1+\vec{p}_2-\vec{k}_3)\prod_{i}d\tau_i\,, \nonumber \\
  &D_2(k_1,k_2,k_3)=\delta^3\big(\sum_i \vec{k}_i \big)~\mu^{-\frac{3 \delta}{2}}  \int d^{d}\vec{p}_1 d^{d}\vec{p}_2P(p_1,p_2,\{k_i\},\{\tau_i\})\sum_{i}n_i\delta\log(-H\tau_i) \nonumber \\ 
   &  \hspace{200pt} \times \delta^{(d)}(\vec{p}_1+\vec{p}_2-\vec{k}_3) \prod_{i}d\tau_i\,,
    \end{align}
    where $D_1$ is the computation of the correlator with $3$-dimensional mode functions and scale factors, and $D_2$ includes the $\mathcal{O}(\delta)$ contribution from them. For $D_1$, computation of the time integrals gives,
    \begin{align}
         D_1(k_1,k_2,k_3)=\delta^3\big(\sum_i \vec{k}_i \big)~\mu^{-\frac{3 \delta}{2}}  \int d^{d}\vec{p}_1 d^{d}\vec{p}_2Q(p_1,p_2,\{k_i\})\delta^{(d)}(\vec{p}_1+\vec{p}_2-\vec{k}_3)\,.
\end{align}
    To compute momentum integrals we use the identity mentioned in \cite{Melville:2021lst} (See Appendix \ref{appendix_1}) and get,
    \begin{align}
          D_1(k_1,k_2,k_3)=\delta^3\big(\sum_i \vec{k}_i \big)~\mu^{-\frac{3 \delta}{2}}  \frac{S_{d-2}}{2}\int_{k_3}^{\infty}dp_{+} \int_{-k_3}^{k_3} dp_{-} \frac{p_1^{d-2}p_2}{k_3}Q(p_1,p_2,\{k_i\}) \,.
    \end{align}
  To simplify the computations further, we turn the integral dimensionless by factoring out powers of one of the momenta (say $k_m$), and work with ratios of momenta 
  \begin{align}
        D_1(k_1,k_2,k_3)=\delta^3\big(\sum_i \vec{k}_i \big)~\mu^{-\frac{3 \delta}{2}}  \frac{S_{d-2}~k^{\delta+N}_m}{2}\int_{1}^{\infty}dq_{+} \int_{-1}^{1} dq_{-} ~{q_1}^{d-2}~q_2~Q(q_1,q_2,\{x_i\}),
  \end{align}
  where $q_i=\frac{p_i}{k_3}$ and $x_i=\frac{k_i}{k_m}$. Computation of the integrals gives the following, 
  \begin{align}
        D_1(k_1,k_2,k_3)=\delta^3\big(\sum_i \vec{k}_i \big)~\mu^{-\frac{3 \delta}{2}}  \frac{S_{d-2}~k^{\delta+N}_m}{2}R(d,\{x_i\}) \,.
  \end{align}
  Finally, we expand everything close to $d=3$ by substituting $d=3+\delta$ and series expanding near $\delta=0$. The function $R(3+\delta,\{x_i\})$ may or may not have a divergence in the limit $\delta\rightarrow 0$. If there is no divergence then we can simply take the limit $\delta\rightarrow 0$ everywhere and get,
  \begin{align}
       D_1(k_1,k_2,k_3)=\delta^3\big(\sum_i \vec{k}_i \big)~\frac{S_{d-2}~k_m^{N}}{2}R(3,\{x_i\}) \,.
  \end{align}
  In the case of a divergence, the function $R(d,\{x_i\})$ will have the form
  \begin{align*}
      R(3+\delta,\{x_i\})=\frac{F_{0}(\{x_i\})}{\delta}+F_1(\{x_i\})+\mathcal{O}(\delta) \,,
  \end{align*} 
  where $F_0$ is the coefficient of the divergent term, and $F_1$ is a function which can have logarithmic dependence on $x_i$. Therefore, we have
  \begin{align}\label{3.26}
        D_1(k_1,k_2,k_3)=\delta^3\big(\sum_i \vec{k}_i \big)~ \frac{S_{d-2}~k_m^{N}}{2}\left\{\frac{F_{0}(\{x_i\})}{\delta}+F_{0}(\{x_i\}) \left( \log{k_m} -\frac{3}{2}\log{\mu} \right)+F_1(\{x_i\})\right\} \,.
  \end{align}
  For the computation of $D_2$, the resulting contribution is equivalent to the logs being pulled outside the integrals as $\log(H/k_T)$, plus some rational functions of momenta (See Appendix~\ref{appendix_3}). Therefore, $D_2$ becomes,
  \begin{align}
      &D_2(k_1,k_2,k_3)=\delta\left[\int d^{d}\vec{p}_1 d^{d}\vec{p}_2P(p_1,p_2,\{k_i\},\{\tau_i\})\delta^{(d)}(\vec{p}_1+\vec{p}_2-\vec{k}_3)\prod_{i}d\tau_i \right] \nonumber \\
      &\hspace{250pt}\times \sum_{i}n_i\log(H/k_T) \delta^3\big(\sum_i \vec{k}_i \big)~ \,, \nonumber \\
     & D_2(k_1,k_2,k_3)=\delta~ D_1\sum_{i}n_i\log(H/k_T) \,. 
  \end{align}
  Combining $D_1$ and $D_2$ gives,
  \begin{align}
      &D=D_1\left(1+\delta\sum_{i}n_i\log(H/k_T)\right)  \,.
      \end{align}
      Therefore, the final expression is given by,
     \begin{tcolorbox}[colback=white,colframe=black]
      \begin{align} \label{3.17}
     & D=\frac{S_{d-2}~k_m^{N}}{2}\left[\frac{F_{0}(\{x_i\})}{\delta}+F_{0}(\{x_i\})\left( \log{k_m} -\frac{3}{2}\log{\mu} +\sum_{i}n_i\log(H/k_T)\right) \right.\nonumber \\
    & \left. +F_1(\{x_i\})\right]~\delta^3\big(\sum_i \vec{k}_i \big) \,.
  \end{align}
  \end{tcolorbox}
If $D_1$ does not have a divergence then $D=D_1$. 

For the loop correction to the power spectrum in \cite{Senatore:2009cf}, $\sum_i n_i =1$. It is clear from the equation above, that unless the number of internal mode functions (i.e. $f_k(\tau)$ at $\tau \neq 0$) and scale factors is such that $\sum_i n_i =1$, the unphysical logs in $D_1$ will not cancel with contributions from $D_2$.  

\subsection{Loop contribution from Diagram I} \label{Dia_1_computation}

Here, we follow the procedure outlined in Sec.~\ref{section:calc_Scheme} to compute the loop contributions coming from diagrams with one quartic and one cubic vertex (Fig.~\ref{Fig:O(c_3 c_4)}). We consider the diagram where $k_3$ contracts with the cubic vertex $\tau_1$. There will be 2 more contributions coming from contractions of $k_1$ and $k_2$ with the cubic vertex, obtained by substituting $k_3 \longleftrightarrow k_1$ and $k_3 \longleftrightarrow k_2$ respectively in the final expressions. 

We first consider the time-ordered diagram Fig.~\ref{Fig:O(c_3 c_4)_1}, with the quartic interaction having only time derivatives, i.e. the $\dot{\pi}^4$ vertex. Using Feynman rules, 
\begin{align}
   & D = - 72~ g_3 g_{4}~\delta^3 \big(\sum_i \vec{k}_i \big) {f}_{k_1}^* {}(0) {f}_{k_2}^* (0) {f}_{k_3}^*(0)\int_{-\infty}^0 d\tau_1 d\tau_2 ~~ \mu^{-\frac{3\delta}{2}} f'_{k_1} (\tau_2) f'_{k_2} (\tau_2) f'_{k_3} (\tau_1)
    \nonumber \\
&a^{d-2}(\tau_1) a^{d-3}(\tau_2)  \int d^d \vec{p}_1 d^d \vec{p}_2 \{ f'_{p_1} (\tau_2) f'{}^*_{p_1} (\tau_1) f'_{p_2} (\tau_2) f'{}^*_{p_2} (\tau_1) \Theta (\tau_1 -\tau_2) \nonumber \\
& + f'_{p_1} (\tau_1) f'{}^*_{p_1} (\tau_2) f'_{p_2} (\tau_1) f'{}^*_{p_2} (\tau_2) \Theta (\tau_2 -\tau_1) \}~ \delta^d (\vec{p}_1 +\vec{p}_2 -\vec{k}_3) \,,
\label{Eqn:modefunc_dia1}
\end{align}
As outlined in Sec. \ref{section:calc_Scheme}, we work with 3d mode functions and ignore $\delta$ dependence from scale factors to compute $D_1(k_1,k_2,k_3)$. We obtain,
\begin{eqnarray}
  D_1 =72~ g_3 g_{4}~\delta^3\big(\sum_i \vec{k}_i \big) \frac{H^9}{\left(-\dot{H} M_{pl}^2 \right)^5} \frac{k_3 (k_1+k_2) (k_3 - 2k_1 -2k_2)}{k_1 k_2 k_T^7} \big(\frac{1}{\delta}+ \frac{1}{2} \log{(k_T k_3)} \nonumber \\
 -\frac{3}{2} \text{log}~\mu \big) + \text{ NLf} \,,
  \label{eqn:Dia1_D1}
\end{eqnarray}
where NLf (Non-logarithmic functions) are simple rational functions of external momenta. Next, we compute $D_2$. As mentioned in Section.~\ref{section:calc_Scheme} (explicit check performed in Appendix~\ref{appendix_2}), $D_2$ can be evaluated  by bringing the log term out of the time integrals as $\log (H /k_T)$, along with some non-logarithmic functions (NLf) of $k_i$. We note that there are 3 internal mode functions and 1 scale factor at $\tau_1$, 4 internal mode functions and 1 scale factor at $\tau_2$. Hence $D_2(k_1,k_2,k_3)$ is given by
\begin{eqnarray}
    D_2 = \delta \left\{\left(\frac{3}{2} -1\right) +\left(\frac{4}{2}-1 \right)\right\} \text{log}\left(\frac{H}{k_T} \right) D_1 +\text{ NLf} \,. \nonumber 
    \end{eqnarray}

Adding this $\mathcal{O}(\delta)$ correction to $D_1$, alongwith contribution from diagrams where $k_1$ and $k_2$ contract with the cubic vertex, gives :
\begin{tcolorbox}[colback=white,colframe=black]
    \begin{align}
   & D =72~ g_3 g_{4}~\delta^3\big(\sum_i \vec{k}_i \big) \frac{H^9}{\left(-\dot{H} M_{pl}^2 \right)^5} \frac{k_3 (k_1+k_2) (k_3 - 2k_1 -2k_2)}{k_1 k_2 k_T^7} \big\{ \frac{1}{\delta}  +\frac{1}{2}\text{log}~\frac{k_3}{k_T}  \nonumber \\
  &-\frac{1}{2}\text{log}~k_T +\frac{3}{2}\text{log}~\frac{H}{\mu}  \big\} + \left\{ k_3 \longleftrightarrow k_1 \right\} + \left\{ k_3 \longleftrightarrow k_2 \right\} +\text{ NLf} \,.
    \label{Eqn:Dia1_timeord_timeder}
\end{align}
\end{tcolorbox}

Next, we compute Fig.~\ref{Fig:O(c_3 c_4)_2} for quartic interactions with temporal derivatives only. Using Feynman rules we get,
\begin{align}
   &D = -72~g_3 g_{4}~\delta^3\big(\sum_i \vec{k}_i \big) {f}_{k_1}^* {}(0) {f}_{k_2}^* (0) {f}_{k_3}(0)\int_{-\infty}^0 d\tau_1 d\tau_2 ~~ \mu^{-\frac{3\delta}{2}} f'_{k_1} (\tau_2) f'_{k_2} (\tau_2) f' {}^*_{k_3} (\tau_1)
    \nonumber \\
&a^{d-2}(\tau_1) a^{d-3}(\tau_2)  \int d^d \vec{p}_1 d^d \vec{p}_2 ~ f'_{p_1} (\tau_2) f'{}^*_{p_1} (\tau_1) f'_{p_2} (\tau_2) f'{}^*_{p_2} (\tau_1) ~ \delta^d (\vec{p}_1 +\vec{p}_2 -\vec{k}_3) \,.
\end{align}

The evaluated answer does not produce any divergences or unphysical logarithms and is given by,
\begin{tcolorbox}[colback=white,colframe=black]
    \begin{align} \label{Eqn:Dia1_D1}
    &D=D_1=72~ g_3 g_{4}~\delta^3\big(\sum_i \vec{k}_i \big) \frac{H^9}{\left(-\dot{H} M_{pl}^2 \right)^5} \frac{k_3 (k_1+k_2) \left(k_T+k_1+k_2 \right) }{2 k_1 k_2 \left(k_3-k_1-k_2\right){}^7} \log \left(\frac{k_T}{2k_3}\right) \nonumber \\
    &+ \left\{ k_3 \longleftrightarrow k_1 \right\} + \left\{ k_3 \longleftrightarrow k_2 \right\} +\text{ NLf} \,.
    \end{align}
\end{tcolorbox}

The apparent folded pole in the above expression actually goes away once we series expand close to the folded limit i.e. $k_3\rightarrow k_1+k_2$ with NLf given by, 
\begin{align}
     & \text{NLf} =72~ g_3 g_{4}~\delta^3\big(\sum_i \vec{k}_i \big) \frac{H^9}{\left(-\dot{H} M_{pl}^2 \right)^5} \frac{1}{8k_1k_2k_3} \frac{1}{120 (k_3 -k_{12})^7 k_T^4} \big( -9 k_3^8+388 k_{12} k_3^7 +1380 k_{12}^2 k_3^6 \nonumber \\
     & +2180 k_{12}^3 k_3^5-190 k_{12}^4 k_3^4-2196 k_{12}^5 k_3^3-1388 k_{12}^6 k_3^2-180 k_{12}^7 k_3+15 k_{12}^8 \nonumber \\
     & -480 k_{12} k_3^7 \log (2)-2880 k_{12}^2 k_3^6 \log (2)-6720 k_{12}^3 k_3^5 \log (2)-7680 k_{12}^4 k_3^4 \log (2) \nonumber \\
     & -4320 k_{12}^5 k_3^3 \log (2)-960 k_{12}^6 k_3^2 \log (2) \big) +...
\end{align}

where the dots represent finite terms in NLf that do not have folded singularity.

Next, we take up quartic interactions with spatial derivatives, i.e. $\dot{\pi}^2 (\partial \pi)^2 $, and compute the time-ordered loop contribution (Fig.~\ref{Fig:O(c_3 c_4)_1}). When both spatial derivatives act on $k_i$ modes (i.e. $f_{k_i}$), the loop contribution is given by 
\begin{align}
& D =18~ g_3^2 ~\delta^3\big(\sum_i \vec{k}_i \big) {f}_{k_1}^* {}(0) {f}_{k_2}^* (0) {f}_{k_3}^*(0) \{\vec{k}_1 \cdot \vec{k}_2\} \int_{-\infty}^0 d\tau_1 d\tau_2 ~~ \mu^{-\frac{3\delta}{2}} a^{d-2}(\tau_1) a^{d-3}(\tau_2)   \nonumber \\ 
& f_{k_1} (\tau_2) f_{k_2} (\tau_2) f'_{k_3} (\tau_1) \int d^d\vec{p}_1 d^d \vec{p}_2 ~ \{ f'_{p_1} (\tau_2) f'{}^*_{p_1} (\tau_1) f'_{p_2} (\tau_2) f'{}^*_{p_2} (\tau_1) \Theta (\tau_1 -\tau_2) \nonumber \\
& + f'_{p_1} (\tau_1) f'{}^*_{p_1} (\tau_2) f'_{p_2} (\tau_1) f'{}^*_{p_2} (\tau_2) \Theta (\tau_2 -\tau_1) \}  \delta^d (\vec{p}_1 +\vec{p}_2 -\vec{k}_3)\,,
\label{Eqn:spatial_der_2_ext}
\end{align}
where $\vec{k}_1 \cdot \vec{k}_2 = -\vec{k}_1 \cdot \left( \vec{k}_1 + \vec{k}_3 \right) = \frac{k_3^2 -k_1^2 -k_2^2}{2}$. 

The spatial derivative can act on a loop mode (i.e. $f_{p_j}$) and on a $k_i$ mode. The contribution will be given by,
\begin{align}
    & D =72~ g_3^2~\delta^3\big(\sum_i \vec{k}_i \big) {f}_{k_1}^* {}(0) {f}_{k_2}^* (0) {f}_{k_3}^*(0) \int_{-\infty}^0 d\tau_1 d\tau_2 ~~ \mu^{-\frac{3\delta}{2}} a^{d-2}(\tau_1) a^{d-3}(\tau_2) \nonumber \\
    & \left\{ f_{k_1} (\tau_2) f'_{k_2} (\tau_2) f'_{k_3} (\tau_1) \vec{k}_1 + f_{k_2}(\tau_2) f'_{k_1}(\tau_2) f'_{k_3} (\tau_1) \vec{k}_2 \right\} \cdot  \int d^d~\vec{p}_1 d^d \vec{p}_2 ~\delta^d (\vec{p}_1 +\vec{p}_2 -\vec{k}_3)
    \nonumber \\ 
  &  \{~ f'{}^*_{p_1} (\tau_1) f'{}^*_{p_2} (\tau_1) \Theta (\tau_1 -\tau_2) \left( f'_{p_1} (\tau_2) f_{p_2} (\tau_2) ~\vec{p}_2 + f_{p_1}(\tau_2) f'_{p_2}(\tau_2) ~\vec{p}_1 \right) \nonumber \\
  &+ f'_{p_1} (\tau_1) f'_{p_2} (\tau_1) \Theta (\tau_2 -\tau_1)  \left( f'{}^*_{p_1} (\tau_2) f_{p_2}^* (\tau_2) ~\vec{p}_2 + f_{p_1}^* (\tau_2) f'{}^*_{p_2} (\tau_2) ~ \vec{p}_1 \right)~\}  \,,
\label{Eqn:spatial_der_1ext1int}
\end{align}

with
\begin{align}
    \vec{k}_1 \cdot \vec{p}_1 = k_1 p_1 \left[ \cos \theta_1 \cos \theta_k + \sin \theta_1 \sin \theta_k \cos \theta_2 \right] \,,  \label{eqn:spatial_Der_mom_identity} 
\end{align}
where $\cos \theta_k$ is angle between $k_1$ and $k_3$, given by $\cos \theta_k = \frac{k_2^2-k_1^2-k_3^2}{2k_1k_3}$. $\theta_1,\theta_2$ are polar angles of $p_1$, with $\cos \theta_1=\frac{k_3^2+p_1^2-p_2^2}{2k_3p_1}$. However we only need to keep the first term in Eq.~\eqref{eqn:spatial_Der_mom_identity} when evaluating the momentum integrals using the identity in Appendix \ref{appendix_1}, since the integrand is odd over $\theta_2 \in (0,\pi)$ due to the $\cos \theta_2$ term. Other dot products can be converted to $\vec{k}_1 \cdot \vec{p}_1 $ using the momentum conserving delta functions.

Finally, the contribution when both spatial derivatives act on loop mode functions is given by,

\begin{align}
    D = 18~ g_3^2~ \delta^3\big(\sum_i \vec{k}_i \big) {f}_{k_1}^* {}(0) {f}_{k_2}^* (0) {f}_{k_3}^*(0) \int_{-\infty}^0 d\tau_1 d\tau_2 ~~ \mu^{-\frac{3\delta}{2}} a^{d-2}(\tau_1) a^{d-3}(\tau_2) \nonumber \\ 
    f'_{k_1} (\tau_2) f'_{k_2} (\tau_2) f'_{k_3} (\tau_1) \int d^d~\vec{p}_1 d^d \vec{p}_2 ~ \{ f_{p_1} (\tau_2) f'{}^*_{p_1} (\tau_1) f_{p_2} (\tau_2) f'{}^*_{p_2} (\tau_1) \Theta (\tau_1 -\tau_2) \nonumber \\
+ f'_{p_1} (\tau_1) f_{p_1}^* (\tau_2) f'_{p_2} (\tau_1) f_{p_2}^* (\tau_2) \Theta (\tau_2 -\tau_1) \} ~\left\{ \vec{p}_1 \cdot \vec{p}_2 \right\}~ \delta^d (\vec{p}_1 +\vec{p}_2 -\vec{k}_3)\,,
\label{Eqn:spatial_der_2_int}
\end{align}

with $\vec{p}_1 \cdot \vec{p}_2 = \vec{p}_1 \cdot \left( \vec{k}_3 -\vec{p}_1 \right)=p_1 k_3 \cos \theta_1 - p_1^2 $.

The $\mathcal{O}(\delta)$ contribution from the mode functions to this computation is given by $D_2 = \frac{3 \delta}{2} \log \left(\frac{H}{k_T} \right) D_1$, obtained from the 3 internal mode functions and 1 scale factor at $\tau_1$, 4 mode functions and 1 scale factor at $\tau_2$ (Appendix \ref{appendix_3}). Computing the integrals above and adding them with $D_2$ gives the following contribution to the 1-loop correction with quartic interactions having spatial derivatives,

\begin{tcolorbox}[enhanced jigsaw,breakable,pad at break =1mm,colback=white,colframe=black]
\begin{align}
     & D =-\frac{3}{40} g_3^2~\delta^3\big(\sum_i \vec{k}_i \big) \frac{H^9}{\left(-\dot{H} M_{pl}^2 \right)^5} \frac{k_3}{k_1^3 k_2^3} \frac{ \vec{k}_1 \cdot \vec{k}_2}{k_T^7} \big(40 k_1^4+280 k_2 k_1^3 +55 k_3 k_1^3+480 k_2^2 k_1^2 \nonumber \\
    &+21 k_3^2 k_1^2 +105 k_2 k_3 k_1^2+280 k_2^3 k_1 +7 k_3^3 k_1 +42 k_2 k_3^2 k_1+105 k_2^2 k_3 k_1+40 k_2^4\nonumber \\
 & +k_3^4+7 k_2 k_3^3 +21 k_2^2 k_3^2 +55 k_2^3 k_3 \big)   \left( \frac{1}{\delta} + \frac{1}{2} \log \left(\frac{k_3}{k_T} \right) - \frac{1}{2} \log k_T  + \frac{3}{2} \log \left( \frac{H}{\mu} \right) \right) \nonumber \\
 &-\frac{3}{40}g_3^2 ~\delta^3 \big(\sum_i \vec{k}_i \big)  \frac{H^9}{\left(-\dot{H} M_{pl}^2 \right)^5} \frac{1}{k_1^3 k_2^3 k_3} \frac{1}{k_T^7}  \big( 20 k_1^8+140 k_2 k_1^7+140 k_3 k_1^7 +200 k_2^2 k_1^6\nonumber \\
    & -152 k_3^2 k_1^6+840 k_2 k_3 k_1^6-140 k_2^3 k_1^5-264 k_3^3 k_1^5  -1904 k_2 k_3^2 k_1^5+560 k_2^2 k_3 k_1^5 \nonumber \\
    & -440 k_2^4 k_1^4 +260 k_3^4 k_1^4+56 k_2 k_3^3 k_1^4-248 k_2^2 k_3^2 k_1^4  -1540 k_2^3 k_3 k_1^4-140 k_2^5 k_1^3 \nonumber \\
    & +124 k_3^5 k_1^3+1764 k_2 k_3^4 k_1^3 -2592 k_2^2 k_3^3 k_1^3+3008 k_2^3 k_3^2 k_1^3 -1540 k_2^4 k_3 k_1^3+200 k_2^6 k_1^2 \nonumber \\
    &-128 k_3^6 k_1^2-896 k_2 k_3^5 k_1^2 +3008 k_2^2 k_3^4 k_1^2-2592 k_2^3 k_3^3 k_1^2 -248 k_2^4 k_3^2 k_1^2+560 k_2^5 k_3 k_1^2\nonumber \\
    & +140 k_2^7 k_1-896 k_2^2 k_3^5 k_1 +1764 k_2^3 k_3^4 k_1+56 k_2^4 k_3^3 k_1 -1904 k_2^5 k_3^2 k_1+840 k_2^6 k_3 k_1+20 k_2^8 \nonumber \\
    & -128 k_2^2 k_3^6+124 k_2^3 k_3^5+260 k_2^4 k_3^4-264 k_2^5 k_3^3  -152 k_2^6 k_3^2+140 k_2^7 k_3 \big) \nonumber \\
    &\left( \frac{1}{\delta} + \frac{1}{2} \log \left( \frac{k_3}{k_T} \right) -\frac{1}{2}\log k_T + \frac{3}{2}\log \left( \frac{H}{\mu}\right) \right) +\frac{9}{4} g_3^2~\delta^3\big(\sum_i \vec{k}_i \big) \frac{H^9}{\left(-\dot{H} M_{pl}^2 \right)^5} \nonumber \\
    & \frac{1}{ k_1 k_2 k_3 k_T^7} \big(k_1^4+4 k_2 k_1^3+7 k_3 k_1^3+6 k_2^2 k_1^2+7 k_3^2 k_1^2+21 k_2 k_3 k_1^2+4 k_2^3 k_1-23 k_3^3 k_1 \nonumber \\ 
    &+14 k_2 k_3^2 k_1+21 k_2^2 k_3 k_1+k_2^4-48 k_3^4-23 k_2 k_3^3+7 k_2^2 k_3^2+7 k_2^3 k_3 \big) \nonumber \\
   &  \left( \frac{1}{\delta} + \frac{1}{2} \log \left( \frac{k_3}{k_T} \right) - \frac{1}{2} \log k_T +\frac{3}{2} \log \left( \frac{H}{\mu} \right) \right)  + \left\{ k_3 \longleftrightarrow k_1 \right\} + \left\{ k_3 \longleftrightarrow k_2 \right\}   +~\text{NLf}\,.
   \label{Eqn:Dia1_timeord_spatialder}
\end{align}
\end{tcolorbox}

Lastly, we compute the unordered diagram Fig~\ref{Fig:O(c_3 c_4)_2} with quartic interaction having spatial derivatives. The correlator when both spatial derivatives act on $k$ mode functions, is given by,
\begin{align}
    & D = 18~ g_3^2 ~\delta^3 \big(\sum_i \vec{k}_i \big)  {f}_{k_1}^* {}(0) {f}_{k_2}^* (0) {f}_{k_3}(0) \{\vec{k}_1 \cdot \vec{k}_2\} \int_{-\infty}^0 d\tau_1 d\tau_2 ~~ \mu^{-\frac{3\delta}{2}} a^{d-2}(\tau_1) a^{d-3}(\tau_2)   \nonumber \\ 
& f_{k_1} (\tau_2) f_{k_2} (\tau_2) f'{}^*_{k_3} (\tau_1) \int d^d\vec{p}_1 d^d \vec{p}_2 ~ f'_{p_1} (\tau_2) f'{}^*_{p_1} (\tau_1) f'_{p_2} (\tau_2) f'{}^*_{p_2} (\tau_1) \delta^d (\vec{p}_1 +\vec{p}_2 -\vec{k}_3)\,.
\end{align}
The correlator when the two spatial derivatives act on a $k$ mode and a loop mode function, is given by
\begin{align}
    & D = 72~ g_3^2~\delta^3  \big(\sum_i \vec{k}_i \big) {f}_{k_1}^* {}(0) {f}_{k_2}^* (0) {f}_{k_3} (0) \int_{-\infty}^0 d\tau_1 d\tau_2 ~~ \mu^{-\frac{3\delta}{2}} a^{d-2}(\tau_1) a^{d-3}(\tau_2) \nonumber \\
    & \left\{ f_{k_1} (\tau_2) f'_{k_2} (\tau_2) f'{}^*_{k_3} (\tau_1) \vec{k}_1 + f_{k_2}(\tau_2) f'_{k_1}(\tau_2) f'{}^*_{k_3} (\tau_1) \vec{k}_2 \right\} \cdot  \int d^d~\vec{p}_1 d^d \vec{p}_2 \nonumber \\ 
  & f'{}^*_{p_1} (\tau_1) f'{}^*_{p_2} (\tau_1) \left( f'_{p_1} (\tau_2) f_{p_2} (\tau_2) ~\vec{p}_2 + f_{p_1}(\tau_2) f'_{p_2}(\tau_2) ~\vec{p}_1 \right) ~\delta^d (\vec{p}_1 +\vec{p}_2 -\vec{k}_3) \,.
\end{align}
When both spatial derivatives act on loop mode functions, we have
\begin{align}
     & D = 18 ~g_3^2~ \delta^3  \big(\sum_i \vec{k}_i \big) {f}_{k_1}^* {}(0) {f}_{k_2}^* (0) {f}_{k_3}(0) \int_{-\infty}^0 d\tau_1 d\tau_2 ~~ \mu^{-\frac{3\delta}{2}} a^{d-2}(\tau_1) a^{d-3}(\tau_2) f'_{k_1} (\tau_2) f'_{k_2} (\tau_2)  \nonumber \\ 
    & f'{}^*_{k_3}(\tau_1) \int d^d~\vec{p}_1 d^d \vec{p}_2 ~ f_{p_1} (\tau_2) f'{}^*_{p_1} (\tau_1) f_{p_2} (\tau_2) f'{}^*_{p_2} (\tau_1) ~\left\{ \vec{p}_1 \cdot \vec{p}_2 \right\}~ \delta^d (\vec{p}_1 +\vec{p}_2 -\vec{k}_3)\,. 
\end{align}
Once again the contribution from this unordered diagram does not produce any divergences or unphysical logarithms, and is given by,

\begin{tcolorbox}[colback=white,colframe=black]
    \begin{align}
   & D=D_1=\frac{3}{20} g_3^2~\delta^3 \big(\sum_i \vec{k}_i \big) \frac{H^9}{\left(-\dot{H} M_{pl}^2 \right)^5} \frac{\{\vec{k}_1 \cdot \vec{k}_2\}}{k_1^3 k_2^3} \frac{k_3}{(k_1+k_2-k_3)^7}  \log \left(\frac{2 k_3}{k_T}\right) \big(40 k_1^4 \nonumber \\
   & +280 k_2 k_1^3-55 k_3 k_1^3+480 k_2^2 k_1^2+21 k_3^2 k_1^2-105 k_2 k_3 k_1^2+280 k_2^3 k_1-7 k_3^3 k_1+42 k_2 k_3^2 k_1 \nonumber \\
   &-105 k_2^2 k_3 k_1+40 k_2^4+k_3^4-7 k_2 k_3^3+21 k_2^2 k_3^2-55 k_2^3 k_3 \big)\nonumber \\
  & +\frac{3}{80} g_3^2~\delta^3 \big(\sum_i \vec{k}_i \big) \frac{H^9}{\left(-\dot{H} M_{pl}^2 \right)^5} \frac{1}{k_1^3 k_2^3 k_3}\frac{1}{(k_1+k_2-k_3)^7}  \log \left(\frac{2k_3}{k_T}\right) \big( 20 k_1^8+140 k_2 k_1^7 \nonumber \\
   & +200 k_2^2 k_1^6-152 k_3^2 k_1^6-840 k_2 k_3 k_1^6-140 k_2^3 k_1^5 +264 k_3^3 k_1^5-1904 k_2 k_3^2 k_1^5-560 k_2^2 k_3 k_1^5 \nonumber \\
 &-440 k_2^4 k_1^4+260 k_3^4 k_1^4-56 k_2 k_3^3 k_1^4-248 k_2^2 k_3^2 k_1^4+1540 k_2^3 k_3 k_1^4-140 k_2^5 k_1^3 -124 k_3^5 k_1^3 \nonumber \\
 & +1764 k_2 k_3^4 k_1^3+2592 k_2^2 k_3^3 k_1^3+3008 k_2^3 k_3^2 k_1^3+1540 k_2^4 k_3 k_1^3+200 k_2^6 k_1^2-128 k_3^6 k_1^2 \nonumber \\
 & +896 k_2 k_3^5 k_1^2+3008 k_2^2 k_3^4 k_1^2+2592 k_2^3 k_3^3 k_1^2-248 k_2^4 k_3^2 k_1^2-560 k_2^5 k_3 k_1^2+140 k_2^7 k_1 \nonumber \\
 & +896 k_2^2 k_3^5 k_1+1764 k_2^3 k_3^4 k_1-56 k_2^4 k_3^3 k_1-1904 k_2^5 k_3^2 k_1-840 k_2^6 k_3 k_1+20 k_2^8-128 k_2^2 k_3^6 \nonumber \\
 & -124 k_2^3 k_3^5+260 k_2^4 k_3^4+264 k_2^5 k_3^3-152 k_2^6 k_3^2-140 k_2^7 k_3 -140 k_3 k_1^7 \big) \nonumber \\
&+ \frac{9}{8} g_3^2~\delta^3 \big(\sum_i \vec{k}_i \big) \frac{H^9}{\left(-\dot{H} M_{pl}^2 \right)^5} \frac{1}{k_1 k_2 k_3} \frac{1}{(k_1+k_2-k_3)^7}  \log \left(\frac{k_3}{k_T}\right) ( -k_1^4-4 k_2 k_1^3+7 k_3 k_1^3 \nonumber \\
& -7 k_3^2 k_1^2+21 k_2 k_3 k_1^2-4 k_2^3 k_1-23 k_3^3 k_1-14 k_2 k_3^2 k_1+21 k_2^2 k_3 k_1-k_2^4+48 k_3^4-23 k_2 k_3^3 \nonumber \\
& -6 k_2^2 k_1^2-7 k_2^2 k_3^2+7 k_2^3 k_3) + \left\{ k_3 \longleftrightarrow k_1 \right\} + \left\{ k_3 \longleftrightarrow k_2 \right\} +~\text{NLf} \,.
\end{align}
\end{tcolorbox}
The apparent folded poles are cancelled by contribution from non-log (NLf) terms.

Eq. \eqref{Eqn:Dia1_timeord_timeder} and Eq. \eqref{Eqn:Dia1_timeord_spatialder} depend on the logarithm of a dimensionful quantity, i.e., $\log k_T$. This is due to the fact that we have inserted a factor of $\mu^{-\frac{3\delta}{2}}$ to keep mass dimensions of couplings unchanged in $d$-dimensions, and in doing so the correlator's mass dimension changes and depends on $\delta$. Factors of $\mu^{\delta}$ are scheme dependent (unphysical), only affecting the constant finite part of the answer, they can therefore be inserted at will (provided one does similar manipulation in the counter-term as well). Hence, if one instead inserts a factor of $\mu^{-\delta}$, such that the mass dimension of the correlator is continued to $d$-dimensions, then the unrenormalized answer will not feature such dimensionful logarithms and instead become $\log{\frac{k_T}{\mu}}$. This is precisely the unphysical logarithm of ratio of co-moving momenta and the renormalisation scale.

As mentioned above, these unphysical logarithms, i.e. $-\frac{1}{2}\log{k_T}$ in Eq. \eqref{Eqn:Dia1_timeord_timeder} and Eq. \eqref{Eqn:Dia1_timeord_spatialder} are present even after adding the $\mathcal{O}(\delta)$ contributions from mode functions. These will cancel once contribution from counter terms is taken into account in Sec~\ref{5}, and the final renormalised result features logarithms of dimensionless quantities.

\subsection{Loop contribution from Diagram II} \label{Dia_2_computation}

As previously stated, we will only give results for the explicit computation of Fig.~\ref{Fig:O(c_3^3)_1}.  There will be 2 more contributions coming from contractions of $k_1$ and $k_2$ with the cubic vertex at $\tau_3$, obtained by substituting $k_3 \longleftrightarrow k_1$ and $k_3 \longleftrightarrow k_2$ respectively in the final expressions. As we will show in Sec.~\ref{5}, the unphysical logarithms cancel diagram-by-diagram in the 2 vertex case, as well as for the 3-vertex diagram we are about to compute. Hence, we expect this cancellation to also occur for each diagram in the 3-cubic vertex case.

The time ordered diagram (Fig.~\ref{Fig:O(c_3^3)_1}) is given by 
\begin{align}
    D = -648i~ g_3^3 ~\delta^3  \big(\sum_i \vec{k}_i \big) f_{k_1}^*(0) f_{k_2}^*(0) f_{k_3}^*(0) \iiint_{-\infty}^0
d\tau_1 d\tau_2 d\tau_3 ~\mu^{-\frac{3\delta}{2}} f'_{k_1} (\tau_1) f'_{k_2} (\tau_1) f'_{k_3} (\tau_3) \nonumber \\
\{a(\tau_1) a(\tau_2) a(\tau_3)\}^{d-2} ~\int d^d~\vec{p}_1 d^d \vec{p}_2 ~\delta^d (\vec{p}_1 +\vec{p}_2 -\vec{k}_3) \nonumber \\
\{ f'_{k_3}(\tau_1) f'{}^*_{k_3}(\tau_2) f'_{p_1}(\tau_2) f'{}^*_{p_1}(\tau_3) f'_{p_2}(\tau_2) f'{}^*_{p_2}(\tau_3) \Theta(\tau_3-\tau_2) \Theta(\tau_2-\tau_1) \nonumber \\
+ f'_{k_3}(\tau_1) f'{}^*_{k_3}(\tau_2) f'{}^*_{p_1}(\tau_2) f'_{p_1}(\tau_3) f'{}^*_{p_2}(\tau_2) f'_{p_2}(\tau_3) \Theta(\tau_2-\tau_1) \Theta(\tau_2-\tau_3) \nonumber \\
+ f'{}^*_{k_3}(\tau_1) f'_{k_3}(\tau_2) f'_{p_1}(\tau_2) f'{}^*_{p_1}(\tau_3) f'_{p_2}(\tau_2) f'{}^*_{p_2}(\tau_3) \Theta(\tau_1-\tau_2) \Theta(\tau_3-\tau_2) \nonumber \\
+ f'{}^*_{k_3}(\tau_1) f'_{k_3}(\tau_2) f'{}^*_{p_1}(\tau_2) f'_{p_1}(\tau_3) f'{}^*_{p_2}(\tau_2) f'_{p_2}(\tau_3) \Theta(\tau_1-\tau_2) \Theta(\tau_2-\tau_3)
\}\,,
\end{align}

    To compute $D_2$, we note $3$ internal mode functions and $1$ scale factor at each of the three time vertices $\tau_1$, $\tau_2$ and $\tau_3$. From arguments given in Sec.~\ref{section:calc_Scheme}, we have $D_2 = \frac{3 \delta}{2} \log \left( \frac{H}{k_T} \right) D_1$. Adding this $\mathcal{O}(\delta)$ contribution, this diagram is computed to be, 
    \begin{tcolorbox}[colback=white,colframe=black]
    \begin{align}
      & D= -648~ g_3^3 ~\delta^3 \big(\sum_i \vec{k}_i \big) \frac{H^9}{\left(-\dot{H} M_{pl}^2 \right)^6} \frac{1}{8 k_1 k_2 k_3} \frac{1}{15  k_T^8} \big( k_1^5+5 k_2 k_1^4+11 k_3 k_1^4+10 k_2^2 k_1^3 \nonumber \\
      & +58 k_3^2 k_1^3+44 k_2 k_3 k_1^3+10 k_2^3 k_1^2+198 k_3^3 k_1^2 +174 k_2 k_3^2 k_1^2 +66 k_2^2 k_3 k_1^2+5 k_2^4 k_1-99 k_3^4 k_1 \nonumber \\ & +396 k_2 k_3^3 k_1+174 k_2^2 k_3^2 k_1 +44 k_2^3 k_3 k_1 +k_2^5 +3 k_3^5  -99 k_2 k_3^4+198 k_2^2 k_3^3 \nonumber \\
      & +58 k_2^3 k_3^2+11 k_2^4 k_3 \big) \left\{ \frac{1}{\delta} + \log \left( \frac{k_3}{k_T} \right) -\frac{1}{2}\log k_T +\frac{3}{2} \log \left( \frac{H}{\mu} \right) \right\} \nonumber \\
        & + 648~ g_3^3~\delta^3  \big(\sum_i \vec{k}_i \big)  \frac{H^9}{\left(-\dot{H} M_{pl}^2 \right)^6} \frac{1}{8 k_1 k_2}\frac{2 k_3^3}{15 k_T^8 \left(k_3-k_1-k_2\right){}^3 } \log \left(\frac{k_T}{k_3}\right) (150 k_1^4\nonumber \\
        &+600 k_2 k_1^3  -195 k_3 k_1^3+900 k_2^2 k_1^2 +133 k_3^2 k_1^2-585 k_2 k_3 k_1^2+600 k_2^3 k_1-25 k_3^3 k_1 \nonumber \\
        &+266 k_2 k_3^2 k_1 -585 k_2^2 k_3 k_1+150 k_2^4+k_3^4-25 k_2 k_3^3+133 k_2^2 k_3^2-195 k_2^3 k_3 ) \nonumber \\
        &+ \left\{ k_3 \longleftrightarrow k_1 \right\} + \left\{ k_3 \longleftrightarrow k_2 \right\}+~\text{NLf}\,.
        \label{Eqn:Dia2_timeord}
    \end{align}
    \end{tcolorbox}

Once again, the apparent folded poles in the second term containing $\log (k_T/k_3)$ cancel from the following contribution from NLf,
    \begin{align}
    & \text{NLf}=648~ g_3^3~\delta^3\big(\sum_i \vec{k}_i \big) \frac{H^9}{\left(-\dot{H} M_{pl}^2 \right)^6} \frac{1}{64 k_1 k_2 k_3 k_T^8} \frac{1}{( 450(k_3-k_{12})^3 )}
        (3 k_3^8 (80 \log (2)-2593) \nonumber \\
       & +96 k_{12} k_3^7 (229-145 \log (2))+24 k_{12}^2 k_3^6 (2380 \log (2)-4173)+8 k_{12}^3 k_3^5 (29681-13380 \log (2)) \nonumber \\
        &+10 k_{12}^4 k_3^4 (5520 \log (2)-19729)+16 k_{12}^5 k_3^3 (2293-840 \log (2))+8 k_{12}^6 k_3^2 (943-840 \log (2)) \nonumber \\
        &+8 k_{12}^7 k_3 (173-240 \log (2))+k_{12}^8 (173-240 \log (2)) ) +... \,,
    \end{align}
with $k_{12}=k_1+k_2$ and the dots denote the finite terms in NLf having no folded poles.

Again, the unphysical logarithms $-\frac{1}{2}\log{k_T}$ in Eq. \eqref{Eqn:Dia2_timeord} is present even after adding the $\mathcal{O}(\delta)$ contributions from mode functions, and it cancels after renormalization in Sec~\ref{5}.
    
    \subsection{Contribution from scaleless loops}  \label{Scaleless_integrals}
  It is well known that all scaleless integrals vanish in dimensional regularization. Here, we show that the contributions from diagrams similar to the ones in Fig.~\ref{Fig:scaleless_1} vanish \cite{Lee:2023jby}. In all these diagrams the loop closes on itself, therefore there is no injection of external kinematics. The bulk-to-bulk propagator for the loop is given by,
 
 \begin{align}
     G_p(\tau,\tau)=\frac{H^{2}}{2p^{3}}\left(1+p^{2}\tau^{2}\right)\,.
 \end{align}
There are no exponentials with energy $p$, which implies that after time integration we only get scale-less integrals of $p$ of the form,
 \begin{align}
     \int p^{\alpha} d^{3}p\,,
 \end{align}
with $k$ dependent coefficients. Such integrals are zero in dim-reg. This can be shown very easily : substituting $p=\lambda p'$ in the above integral in $d$ dimensions, we get,
 \begin{align}
     \int  \lambda^{d+\alpha} p'^{\alpha} d^{d}p'\,,
 \end{align}
 and we have,
 \begin{align}
       \int p^{\alpha} d^{d}p= \lambda^{d+\alpha}\int   p'^{\alpha} d^{d}p'\,.
 \end{align}
 The above equality follows because the limits are from $-\infty$ to $\infty$. Now, since $d+\alpha \neq 0$, the only possibility is that the integral is zero. Therefore, all 1-loop diagrams where the loop closes on itself are zero. However, one needs to be careful about diagrams which have a tadpole attached (e.g. Fig.~\ref{Fig:scaleless_2}). Since by translational invariance the tadpole leg has zero momentum therefore in some cases the integrand for the diagram may diverge. This does not happen in our case since every field operator in the interaction Hamiltonian comes with a derivative. Therefore, either the intergand of such diagrams is  zero or finite. The finite case evaluates to zero via the above-mentioned scaleless loop argument.

To see this, note that the propagator arises due to contraction of the tadpole leg coming from the cubic ($\dot \pi^3$) vertex with a mode function coming from either the cubic or a quartic ($\dot \pi^4$ , $\dot \pi ^2 (\partial \pi)^2$) vertex. When mode functions with time derivatives contract to give the propagator, the integrand is proportional to the tadpole leg momentum, $k$ and hence vanishes for $k=0$. When modes with a time and a spatial derivative contract, the propagator is non-vanishing but finite for $k=0$, hence the integrand is finite and the tadpole contribution vanishes by the scaleless arguments given in this section.
    
\section{Renormalisation: cancelling the unphysical logarithms} \label{5}
It is well known that Effective Field theories can be renormalised like usual renormalisable theories provided we restrict to a particular order in the EFT expansion. In a renormalisable field theory ($\mathcal{L}^{(d\leq 4)}$), to renormalise any diagram one does not need to introduce any higher dimensional operators, all divergences can be absorbed in a finite set of parameters present in the lagrangian itself. On the other hand, for an EFT, one needs to include higher dimensional counter-term operators to renormalise the theory, but thankfully only a finite number of counter-terms are needed at a particular order. For instance, a loop diagram computed with two insertions of a dimension six operator is of order $\left(\frac{E}{\Lambda}\right)^{4}$ and therefore, it is necessary to have an eight-dimensional counter operator to renormalise it. \par
In this paper, we computed the three-point function at 1-loop for the following EFT,
\begin{align}
    S= \int d^{4}x \sqrt{-g}\bigg[\left(\dot{\pi_c}^{2}-\frac{(\partial\pi_c)^{2}}{a^{2}}\right)+\frac{\tilde{g}_3}{\Lambda^{2}}\dot{\pi_c}^{3} + \frac{\tilde{g}_{4}}{\Lambda^{4}}\dot{\pi_c}^{4}-\frac{3 \tilde{g}_3}{2\Lambda^{4}}\dot{\pi_c}^{2}\left(\frac{\partial\pi_c}{a}\right)^{2}\bigg]\,,
\end{align}
where we have now written the EFT lagrangian given in Eq. \eqref{EFToI} in terms of inverse powers of some UV cutoff scale $\Lambda$. This makes the power counting manifest. We have assumed that all operators are generated by integrating out the same scale in the UV. In the expression above $\{\tilde{g_i}\}$ are all dimensionless and $\pi_c=\sqrt{-\dot{H}}M_{pl}\pi$ (canonically normalised). The EFToI Lagrangian was written in a similar form also in \cite{Baumann:2015nta}. It is now clear that our loop results are of order $\left(\frac{H}{\Lambda}\right)^{6}$, therefore we need ten-dimensional cubic counter term operators to renormalise it. We do not explicitly write down all possible operators. Instead, we show the following: given that the counter-terms, which give the same kinematical structures as that of terms multiplying the divergence, exist then once added to the bare answers they get rid of the unphysical logarithms. The most important takeaway from this section is that the unphysical structures of the form $\log{\frac{k}{\mu}}$ we got in the previous sections are corrected after renormalisation and we finally get logarithms of ratios of comoving momenta. The expected structure of counter-term operators is the following,
\begin{align}
    \mathcal{O}^{\text{CT}}\sim \frac{1}{\delta}\int \mu^{-\delta/2} a^{-3+\delta} \partial^{7}\left(\pi^{3}\right)\,,
\end{align}
where $\partial^{7}$ are seven space/time derivatives. The three internal mode functions plus the scale factors bring the following $\delta$ dependence in the in-in integrals,
\begin{align}
     \frac{1}{\delta} \int d\tau \mu^{-\delta/2} (-H\tau)^{\delta/2}...~\,,
\end{align}
where the dots represent the remaining terms in the in-in integral. This expression can be simplified to yield,
\begin{align}
     \int d\tau\left(\frac{1}{\delta}-\frac{1}{2} \log{\mu} +\frac{1}{2}\log{(-H\tau)}\right)...~\,.
\end{align}
Once again we can pull $\log{(-H\tau)}$ out of the integral, as we have done before, as $\log{\left(\frac{H}{k_T}\right)}$. Hence, we factor out $\left(\frac{1}{\delta}+\frac{1}{2}\log{\left(\frac{H}{\mu}\right)} -\frac{1}{2}\log{k_T}\right)$ from all the counter terms. This implies that the coefficient of the unphysical logarithm and the divergent piece have a relative factor of $-1/2$, which is exactly the case for the unrenormalised answer of each diagram. Therefore if the counter-term is set appropriately to cancel the divergence, then it will automatically get rid of the unphysical logarithm. As mentioned above we do not explicitly renormalise the theory. Instead we assume such counterterms exist, i.e. we can get the kinematic structures that multiply the divergent terms in Sec. \ref{4}, ensuring that the divergence can be canceled by the counter terms. More explicitly, we saw that for diagrams computed in Sec. \ref{4}, 
\begin{align}\label{4.4}
 D(k_1,k_2,k_3)=\frac{S_{d-2}k_3^{N}}{2} F_{0}(\{x_i\}) \times \left(\frac{1}{\delta}+\frac{1}{2} \log{\left(\frac{k_3}{k_T}\right)}-\frac{1}{2}\log{k_T}+\frac{3}{2} \log{\left(\frac{H}{\mu}\right)}+F_1(\{x_i\})\right) \nonumber \\
  +\left\{ k_3 \longleftrightarrow k_1 \right\} + \left\{ k_3 \longleftrightarrow k_2 \right\}\,.
\end{align}
Now, we assume that a linear combination of counter-term operators with an appropriate choice of coefficients can yield an answer of the form,
\begin{align}\label{4.5}
    D^{\text{CT}}=-\left[\frac{S_{d-2}k_3^{N}}{2}F_{0}(\{x_i\}) \times \left(\frac{1}{\delta}+\frac{1}{2}\log{\left(\frac{H}{\mu}\right)} -\frac{1}{2} \log{k_T}\right)+ \left\{ k_3 \longleftrightarrow k_1 \right\} + \left\{ k_3 \longleftrightarrow k_2 \right\}\right]\,.
\end{align}
Adding Eq. \eqref{4.4} and Eq. \eqref{4.5} we get,
\begin{tcolorbox}[colback=white,colframe=black]\begin{align}
    D^{\text{renorm}}=\delta^3 \big(\sum_i \vec{k}_i \big) \frac{S_{d-2}k_3^{N}}{2}\left(F_{0}(\{x_i\})\log{\left(\frac{k_3}{k_T}\right)}+F_{0}(\{x_i\})\log{\left(\frac{H}{\mu}\right)}+F_1(\{x_i\})\right) \nonumber \\
    +\left\{ k_3 \longleftrightarrow k_1 \right\} + \left\{ k_3 \longleftrightarrow k_2 \right\}\,.
\end{align}\end{tcolorbox}
This gets rid of unphysical logarithms of co-moving momenta in all the diagrams computed. 

It is important to note that such a feature of log cancellation from counter terms was missing from the two-point function calculation performed in \cite{Senatore:2009cf}. The reason is pretty straightforward, in this case, there are two scale factors and six mode functions and this is just right to cancel the unphysical logs. Let us review this very quickly, the  $\delta$ dependence of the loop integral is,
\begin{align}
    \mu^{-\delta/2}\int d\tau d^{3+\delta}p_1d^{3+\delta}p_2(-H\tau)^{\delta}.....\,,
\end{align}
Again the logarithm can be factored out at scale $k$ which is the only scale in the problem and the integral evaluates to the form given in Eq. \eqref{3.17} with $\sum_i n_i=1$, and we get, 
\begin{align}
     D&=\frac{\mu^{-\delta/2}S_{d-2}k^{N}}{2}\left(\frac{F_{0}}{\delta}+F_{0}\log{k}+F_{0}\log(H/{k})+\text{finite}\right) \nonumber \\
      &=\frac{S_{d-2}k^{N}}{2}\left(\frac{F_{0}}{\delta}+F_{0}\log{\frac{H}{\mu}}+\text{finite}\right) \,.
\end{align}
So in this case the unphysical logarithms cancel without renormalising. We believe that for general one-loop diagrams, the unphysical logarithms will be canceled only after renormalizing the results unless there is some cancellation as in the two-point case.

\section{Cutoff Regularisation: a cross check} \label{Sec.cutoff}

If the full mode functions are used in our computations instead of the Taylor expanded ones, evaluation of the time integrals results in complicated hypergeometric functions, and the problem becomes intractable. Thus we work with the modes expanded about $\delta=0$. Now, the expansion in $\delta$ for $(-H \tau)^{\delta}=\exp{\delta \log{(-H \tau)}}$ converges for all values of $\delta$. Also, we only had to evaluate the $\mathcal{O}(\delta)$ term, since the higher order terms did not produce higher order poles upon evaluation of the loop integrals, hence vanishing when $\delta \rightarrow 0$.

Note that in our computations, we had to commute the summation operator across the integrals. There is a subtlety involved here. For the loop integrals to converge, $\delta$ must be analytically continued to some point in the complex plane, and arguing the convergence of the sum of integrals is not straightforward at this value of $\delta$. This makes the commutation of summation and integration operations tricky. However, to demonstrate that our method of implementing the dimensional regularisation still computes the logarithmic running reliably, we compute one of the diagrams in the cutoff-regularisation, as follows.

We focus on the time ordered contribution to Diagram I coming from the interactions with only time derivatives i.e., $\dot{\pi}^{3}$, $\dot{\pi}^{4}$. Note that there is another contribution to this diagram coming from the quartic interaction $\dot{\pi}^2 (\partial \pi)^2$, but since it has a different coupling constant, this contribution cannot alter the log structure we compute below. It is interesting to note that in cutoff regularisation, the mode functions are 3-dimensional, hence the contribution from the counter terms, which are tree-level diagrams, will never produce $\text{log}~k$ terms. Thus we expect the unrenormalised answer not to feature any unphysical logarithms, which turns out to be true in the following. 

To evaluate Eqn.~\eqref{Eqn:modefunc_dia1} in cutoff regularisation, we regulate the momentum integrals by placing a cutoff at $\Lambda a(\tau_f)$, where $\Lambda$ is the physical cutoff and $\tau_f$ is the time being integrated first, e.g. for the $\Theta(\tau_1 -\tau_2)$ part of the integrand $\tau_f =\tau_2$. We integrate the loop integrals first, since they are now $\tau$ dependent, using the following identity :

\begin{align}
    \int d^3 \vec{p}_1 d^3 \vec{p}_2 Q(p_1,p_2,{k_i}) \delta^3 (\vec{p}_1 + \vec{p}_2 +\vec{k}_3)=2 \pi \int_0^{\Lambda a(\tau_f)} dp_1 \int_{|\vec{p}_1-\vec{k}_3|}^{\vec{p}_1 +\vec{k}_3} dp_2 \frac{p_1 p_2}{k_3} Q(p_1,p_2,{k_i}) \,,
\end{align}

Evaluating the $\tau_1 >\tau_2$ piece of the loop integrals gives the following :

\begin{align}
   & D=  - 36 ~ g_3 g_{4}~\delta^3 \big(\sum_i \vec{k}_i \big)  \int_{-\infty}^0 d\tau_1 \int_{-\infty}^{\tau_1 \left(1+ \frac{H}{\Lambda} \right)} d\tau_2 ~~ \{ \frac{1}{16 k_3 \left(\tau _1-\tau _2\right)^6} \tau _1^2 \tau _2^4 \big(e^{i \left(k_1+k_2+k_3\right) \tau _2} \nonumber \\
   & \left(-k_3^2 \left(\tau _1-\tau _2\right)^2+5 i k_3 \left(\tau _1-\tau _2\right)  +8 \right)  +e^{i \left(k_1+k_2+3 k_3\right) \tau _2-2 i k_3 \tau _1} (8 k_3^4 \left(\tau _1-\tau _2\right)^4 \nonumber \\
   &-20 i k_3^3 \left(\tau _1-\tau _2\right)^3+k_3^2 \left(\tau _1-\tau _2\right)^2 (-27 +8 e^{2 i k_3 \left(\tau _1-\tau _2\right)} ) -i k_3 \left(\tau _1-\tau _2\right) \left(-21+16 e^{2 i k_3 \left(\tau _1-\tau _2\right)}\right) \nonumber \\
   & -16 e^{2 i k_3 \left(\tau _1-\tau _2\right)}+8 )\big) + ... \nonumber \\
   &+ \frac{1}{240 k_3 \left(\tau _1-\tau _2\right){}^6}\tau _1^2 \tau _2^4 (-2 i k_3^5 \left(\tau _1-\tau _2\right)^5 e^{2 i k_3 \left(\tau _1-\tau _2\right)}-10 k_3^4 \left(\tau _1-\tau _2\right)^4 \left(12+e^{2 i k_3 \left(\tau _1-\tau _2\right)}\right) \nonumber \\
   & +20 i k_3^3 \left(\tau _1-\tau _2\right){}^3 \left(15+2 e^{2 i k_3 \left(\tau _1-\tau _2\right)}\right)-15 k_3^2 \left(\tau _1-\tau _2\right){}^2 \left(-27+e^{2 i k_3 \left(\tau _1-\tau _2\right)}\right)+15 i k_3 \nonumber \\
   & \left(\tau _1-\tau _2\right) \left(-21+5 e^{2 i k_3 \left(\tau _1-\tau _2\right)}\right)+120 \left(-1+e^{2 i k_3 \left(\tau _1-\tau _2\right)}\right) ) e^{\left(i \left(k_1+k_2\right) \tau _2-i k_3 \left(2 \tau _1-3 \tau _2\right)\right)} \} \,,
   \label{Eqn:t1>t2}
\end{align}
where we regulate the time integrals with the same cutoff $\Lambda$. The first $4$ lines correspond to $q_1>k_3$, the last $3$ lines correspond to $q_1<k_3$. Similarly, evaluating the $\tau_2>\tau_1$ piece gives,
\begin{align}
    & D=  - 36 ~ g_3 g_{4}~\delta^3 \big(\sum_i \vec{k}_i \big)  \int_{-\infty}^0 d\tau_2 \int_{-\infty}^{\tau_2 \left(1+ \frac{H}{\Lambda} \right)} d\tau_1 ~~ \{ \frac{1}{16 k_3 \left(\tau _1-\tau _2\right){}^6} \tau _1^2 \tau_2^4 e^{i \left(2 k_3 \tau _1+\left(k_1+k_2-3 k_3\right) \tau _2\right)} \nonumber \\
    & (8 k_3^4 \left(\tau _1-\tau _2\right){}^4 e^{2 i k_3 \tau _1}+20 i k_3^3 \left(\tau _1-\tau _2\right){}^3 e^{2 i k_3 \tau _1}-k_3^2 \left(\tau _1-\tau _2\right){}^2 \left(27 e^{2 i k_3 \tau _1}-7 e^{2 i k_3 \tau _2}\right) \nonumber \\
    & -i k_3 \left(\tau _1-\tau _2\right) \left(21 e^{2 i k_3 \tau _1}-11 e^{2 i k_3 \tau _2}\right)+8 \left(e^{2 i k_3 \tau _1}-e^{2 i k_3 \tau _2}\right) ) +... \nonumber \\ 
    & + \frac{1}{4 k_3 \left(\tau _1-\tau _2\right){}^3} \tau _1^2 \tau _2^4 e^{i \left(k_3 \left(2 \tau _1-\tau _2\right)+\left(k_1+k_2\right) \tau _2 \right)}  (\frac{1}{5} i k_3^5 \left(\tau _1-\tau _2\right){}^2 e^{i \left(k_3 \left(2 \tau _1-\tau _2\right)+\left(k_1+k_2\right) \tau _2\right) } \nonumber \\ 
    & +\frac{1}{2} k_3^4 \left(\tau _1-\tau _2\right) \left(1-i k_3 \left(\tau _1-\tau _2\right)\right) e^{i \left(k_3 \left(2 \tau _1-\tau _2\right)+\left(k_1+k_2\right) \tau _2\right)} +\frac{1}{3} i k_3^3 (k_3^2 \left(\tau _1-\tau _2\right){}^2 \nonumber \\ 
    & +2 i k_3 \left(\tau _1-\tau _2\right)-2 ) -\frac{1}{4 \left(\tau _1-\tau _2\right){}^3} e^{i \left(2 k_3 \tau _1+\left(k_1+k_2-k_3\right) \tau _2\right)} \left(k_3^2 \left(\tau _1-\tau _2\right){}^2+5 i k_3 \left(\tau _1-\tau _2\right)-8\right)  \nonumber \\
    & +\frac{1}{4 \left(\tau _1-\tau _2\right){}^3} e^{i \left(4 k_3 \tau _1+\left(k_1+k_2-3 k_3\right) \tau _2\right)} (-8 k_3^4 \left(\tau _1-\tau _2\right){}^4-20 i k_3^3 \left(\tau _1-\tau _2\right){}^3 \nonumber \\
    & +27 k_3^2 \left(\tau _1-\tau _2\right){}^2+21 i k_3 \left(\tau _1-\tau _2\right)-8 ) ) \}  \,,
    \label{Eqn:t2>t1}
\end{align}
 where the first $3$ lines correspond to $q_1>k_3$ and the last $5$ lines correspond to $q_1<k_3$. In Eqns.~\eqref{Eqn:t1>t2} and \eqref{Eqn:t2>t1} the dots represent contributions that are proportional to the term ($\text{Exp} \big\{ \frac{\Lambda}{H} \big(\frac{\tau_s}{\tau_f}-1 \big) \big\}$), where the $\Theta$ function imposes $\tau_s > \tau_f$. These terms do not produce any logarithms (as was also noted in \cite{Senatore:2009cf}).

Evaluating the time integrals and adding all contributions, we arrive at the following answer,

\begin{align}
   & D =72~ g_3 g_{4}~\delta^3\big(\sum_i \vec{k}_i \big) \frac{H^9}{\left(-\dot{H} M_{pl}^2 \right)^5} \frac{k_3 (k_1+k_2) (k_3 - 2k_1 -2k_2)}{k_1 k_2 k_T^7} \big\{ \text{log}~ \frac{H}{\Lambda}  \nonumber \\
  & + \frac{1}{2} \text{log}~\frac{k_3}{k_T} \big\} + \left\{ k_3 \longleftrightarrow k_1 \right\} + \left\{ k_3 \longleftrightarrow k_2 \right\} +\text{L} \,.
\end{align}
where $\text{L}$ denotes power divergences in $\Lambda$ and other non-log finite terms.

After renormalisation $\log{\frac{H}{\Lambda}}\longrightarrow \log{\frac{H}{\mu}}$. Therefore the logarithm running, as well as the structure of the coefficients multiplying them, matches with the results of our previous dim reg computation after adding contribution from counter terms. This shows that the procedure of expanding in $\delta$ gives the correct log running. 

\section{Conclusion} \label{6}
In this work, we computed the bispectrum (3-point function) at one loop in the decoupling limit of EFToI. Our main objective was to explore the possible analytical structures at one loop. We show that the loop correction depends on rational functions of external momenta, alongwith structures of the form $\log{\left(H / \mu\right)}$ and $\log{\left(k_i / k_T\right)}$. Interestingly, we note that unlike the 1-loop correction to two-point function \cite{Senatore:2009cf}, the unphysical logarithms of co-moving momenta in our results do not cancel upon analytically continuing the mode functions to $d$-dimensions. They are removed only after our results are renormalized. We believe such cancellations of unphysical logs will happen for general one-loop diagrams only after renormalizing the results, unless there is some cancellation as in the two-point case.

Note that we did not explicitly renormalise our results but simply showed that once an appropriate combination of counter-terms is chosen to cancel the divergence, it will always automatically get rid of the unphysical logarithms. \par

\acknowledgments
We thank Guilherme Leite Pimentel for valuable comments. We would also like to thank Arhum Ansari, K S Dhruva and Saurabh Pant for useful discussions. SB acknowledges support through the PMRF Fellowship of the Gov. of India.  
\appendix

\section{Loop Momentum Identity}

Here, we compute momentum integrals given by
\begin{align}
    D_1(k_i)= \int d^d \vec{p}_1 d^d \vec{p}_2 ~Q(p_1,p_2,\{ k_i \}) \delta^d (\vec{p}_1 +\vec{p}_2 -\vec{k}_3) \,.
    \label{eqn:mom_identity}
\end{align}
To evaluate this, we note that Eq.~\eqref{eqn:mom_identity} can be rewritten, by carrying out the integral over $\vec{p}_2$ using the delta function, as follows
\begin{align}
   & = \int d^d \vec{p}_1 ~Q(p_1,|\vec{k}_3-\vec{p}_1|,\{k_i\}) =\int p_1 ^{d-1} Q(p_1,|\vec{k}_3-\vec{p}_1|,\{k_i\})~ dp_1 d\Omega_d \,, \nonumber \\
   & =\int p_1 ^{d-1} ~Q(p_1,|\vec{k}_3-\vec{p}_1|,\{k_i\})~ dp_1 \sin^{d-2}{\theta_1} \sin^{d-3}{\theta_2}...\sin{\theta_{d-2}} ~ d\theta_1 d\theta_2 d\theta_{d-2} d\phi \,,
    \label{eqn:mom_identity_2}
\end{align}

where $\phi$ is the azimuthal angle and $\theta_i$'s denote polar angles. If $\hat{k}_3$ is chosen along axis with respect to which $\theta_1$ is measured, then
\begin{align*}
    & |\vec{k}_3-\vec{p}_1| =\sqrt{k_3^2 +p_1^2-2k_3 p_1 \cos{\theta_1}}=p_2 \\
   & \text{or,}~\frac{k p_1}{p_2} \sin{\theta_1} d\theta_1 =dp_2
\end{align*}

Thus, a variable change from $\theta_1$ to $p_2$ in Eq. \eqref{eqn:mom_identity_2} gives us 
\begin{align}
    =\int \frac{p_1 ^{d-2} p_2}{k_3} ~Q(p_1,p_2,\{k_i\}) \sin^{d-3}\theta_1 ~dp_1 dp_2 d\Omega_{d-1} \,, \nonumber \\
    =S_{d-2} \int \frac{p_1 ^{d-2} p_2}{k_3} ~Q(p_1,p_2,\{k_i\}) \sin^{d-3}\theta_1 ~dp_1 dp_2 \,,
\end{align}

where $S_{d-2}$ is surface area of a $(d-2)$ - unit sphere, and $\sin{\theta_1}=\sqrt{1-\left( \frac{k_3^2 +p_1^2-p_2^2}{2k_3 p_1}\right)^2}$ . Defining $p_+=p_1+p_2$ and $p_-=p_1-p_2$, we get

\begin{align}
    &\int d^d \vec{p}_1 d^d \vec{p}_2 ~Q(p_1,p_2,\{k_i\}) \delta^d (\vec{p}_1 +\vec{p}_2 -\vec{k}_3) \nonumber \\
    & = \frac{S_{d-2}}{2} \int_{k_3}^{\infty} dp_+ \int_{-k_3}^{k_3} dp_- \frac{p_1 ^{d-2} p_2}{k_3} \sin^{d-3}{\theta_1}~Q(p_1,p_2,\{k_i\}) \,.
\end{align}

\noindent The factor of $\sin^{d-3}\theta_1$  was dropped in \cite{Melville:2021lst}. While it does not contribute to the $\mathcal{O}(\frac{1}{\delta})$ divergent terms in the correlator, it does give finite contributions and hence is important in determining the shape and pole structure of the correlator away from the equilateral limit. 

\label{appendix_1}

\section{Computation of $D_2$ for Diagram I}

\label{appendix_2}

Here we explicitly compute the $\mathcal{O}(\delta)$ contribution from mode functions and scale factors for the diagram with time ordered vertices in Fig.~\ref{Fig:O(c_3 c_4)}. The integral is given by 
\begin{align}
    D_2=g_3 g_{4}~\delta^3\big(\sum_i \vec{k}_i \big) \frac{H^9}{\left(-\dot{H} M_{pl}^2 \right)^5} \frac{1}{8k_1k_2k_3} \int d^d\vec{p}_1 d^d\vec{p}_2 ~\frac{p_1 p_2}{4} \delta  (I_1 +I_2) \,,
\end{align}
where $I_1$ and $I_2$ are the integrals with time orderings $\tau_1>\tau_2$ and $\tau_2>\tau_1$ respectively,
\begin{align}
    & I_1 (k_i,p_j) = \int_{-\infty}^{0} d\tau_1 \int _{-\infty}^{\tau_1} d\tau_2 ~e^{i(k_{12}+p_+)\tau_2}e^{i(k_3-p_+)\tau_1} \left( \frac{1}{2} \text{log}(-H \tau_1) +\text{log} (-H \tau_2) \right) \nonumber \\
    & \hspace{40pt}= I_1^{\tau_1}(k_i,p_j) + I_1^{\tau_2}(k_i,p_j) \,, \nonumber \\
    & I_2(k_i,p_j)= \int_{-\infty}^{0} d\tau_2 \int _{-\infty}^{\tau_2} d\tau_1 ~e^{i(k_{12}-p_+)\tau_2}e^{i(k_3+p_+)\tau_1} \left( \frac{1}{2} \text{log}(-H \tau_1) +\text{log} (-H \tau_2) \right) \nonumber \\
    &\hspace{40pt}=I_2^{\tau_1}(k_i,p_j) + I_2^{\tau_2}(k_i,p_j) \,,
\end{align}

where we refer to the terms having log dependence on the time integrated from $-\infty$ to $0$ as $I_1^{\tau_1}(k_i,p_j)$ and $I_2^{\tau_2}(k_i,p_j)$, i.e., $\text{log}(-H \tau_1)$ in $I_1$ and $\text{log}(-H \tau_2)$ in $I_2$ respectively. Next, we consider terms having log dependence on the time being integrated from $-\infty$ to $\tau_{1(2)}$, and call them $I_1^{\tau_2}(k_i,p_j)$ and $I_2^{\tau_1}(k_i,p_j)$, i.e., $\text{log}(-H \tau_2)$ in $I_1$ and $\text{log}(-H \tau_1)$ in $I_2$ respectively.
\begin{itemize}
    \item From $I_1^{\tau_1}$ and $I_2^{\tau_2}$, we get functions of $p_+,k_{12}$ and $k_3$, plus terms dependent on $\log k_T$ . The former gives a $\frac{1}{\delta}$ divergent term after the loop integrals, independent of logarithms, hence giving a finite contribution to $D_2$. The log term gives the following finite contribution to $D_2$ after performing loop integrals,
    \begin{align}
        g_3 g_{4}~\delta^3 \big(\sum_i \vec{k}_i \big) \frac{H^9}{\left(-\dot{H} M_{pl}^2 \right)^5} \frac{k_3 (k_1+k_2) (k_3 - 2k_1 -2k_2)}{k_1 k_2 k_T^7} \frac{3}{2} \log \left(\frac{H}{k_T}\right) \,.
        \label{eqn:log_D2}
    \end{align}
 \item From $I_1^{\tau_2}$ and $I_2^{\tau_1}$, we once again get functions of $q_+,k_{12}$ and $k_3$, and a $\text{log}~k_T$ term. Additionally, one also gets terms dependent on $\text{log}(p_++k_{12})$ and $\text{log}(p_++k_{3})$ for $I_1^{\tau_2}$ and $I_2^{\tau_1}$ respectively. However, upon $p_-$ integration, the $p_+$ dependence of the result
 \begin{align}
 \sim  \frac{p_+^4 \left(p_+^2-k_{12}^2\right){}^{\delta /2} \log \left(p_++\frac{k_{12}}{k_3}\right)}{\left(p_+-k_{12}\right){}^3 \left(p_++\frac{k_{12}}{k_3}\right){}^5} \,,
 \label{eqn:conv_q+}
 \end{align}
 vanishes in the limit $p_+ \rightarrow \infty$. Thus the contributions from such terms to $D_2$ are of $\mathcal{O}({\delta})$, and vanish when $d\rightarrow 3$.
 
 The remaining log independent terms give a finite, log-free contribution to $D_2$, and the contribution from $\text{log}~k_T$ terms after performing loop integrals is the same as Eq. \eqref{eqn:log_D2}
    \end{itemize}
Hence, summing the contributions, we have
\begin{align}
    D_2=g_3 g_{4}~\delta^3\big(\sum_i \vec{k}_i \big) \frac{H^9}{\left(-\dot{H} M_{pl}^2 \right)^5} \frac{k_3 (k_1+k_2) (k_3 - 2k_1 -2k_2)}{k_1 k_2 k_T^7} 3 \log \left(\frac{H}{k_T}\right) + \text{finite} \,.
\end{align}
Adding this to $D_1$ gives Eq. \eqref{Eqn:Dia1_timeord_timeder}

 \section{Computation of $\mathcal{O}(\delta)$ terms containing $\log \left(-H \tau\right)$}
 \label{appendix_3}
 
 In Appendix~\ref{appendix_2}, we showed explicitly that $D_2$ can be computed by pulling the log term out of the integrals and substituting $\tau \rightarrow 1/k_T$, along with some finite terms.  In this appendix, we show that this is also true for the $D_2$ contributions for all diagrams computed in Sec.~\ref{4}
 
 Note that $D_2$ is given by the same integral as $D_1$ with an additional factor of $\log \tau$. To compute this contribution, one would need to evaluate up to three time integrals, with logarithms dependent on any of the time vertices. The explicit computation of $D_2$ will be lengthy since each such integral can have various time orderings. Instead, let us take a look at the general structure of these integrals. 

 For diagrams with a cubic and a quartic interaction, the time integrals have the form,
 
 \begin{align}
      \int_{-\infty}^0 d\tau_1 \int_{-\infty}^{\tau_1} d\tau_2 ~\tau_1^n \tau_2^m e^{i a \tau_1} e^{i b \tau_2} \log \left(-H \tau_{1(2)} \right)  \,,
      \label{eqn:2vertices}
 \end{align}
 where $a+b=k_T$. In the integrals we compute, $(m,n)$ take the values $(2,2)$, $(2,3)$ in Eqns.~\eqref{Eqn:spatial_der_1ext1int},~\eqref{Eqn:spatial_der_2_ext},~\eqref{Eqn:spatial_der_2_int}, and $(2,4)$ in Eqns.~\eqref{Eqn:modefunc_dia1},~\eqref{Eqn:spatial_der_1ext1int},~\eqref{Eqn:spatial_der_2_ext},~\eqref{Eqn:spatial_der_2_int}. The log acts on $\tau_1$ or $\tau_2$ . For three cubic diagrams, the time integrals are of the type
 \begin{align}
      \int_{-\infty}^0 d\tau_1 \int_{-\infty}^{\tau_1} d\tau_2 \int_{-\infty}^{\tau_2} d\tau_3 ~\tau_1^m \tau_2^n \tau_3^{l} e^{i a \tau_1} e^{i b \tau_2} e^{i c \tau_3} \log \left( -H\tau_{1(2,3)} \right) \,,
      \label{eqn:3vertices}
 \end{align}
 where $a+b+c=k_T$ and $m,n,l=2$ for our computations. The log acts on $\tau_1$,$\tau_2$ or $\tau_3$.

\begin{itemize}

    \item \textbf{Integrals with $\mathbf{\log\left(-H \tau_1\right)}$ :} Performing the integral over $\tau_2$ for Eq. \eqref{eqn:2vertices} (and over $\tau_3$ for Eq. \eqref{eqn:3vertices}), we are left with 
    \begin{align}
        \int_{-\infty}^0 d\tau_1 ~\tau_1^m e^{i k_T \tau_1} g(k_i,p_j,\tau_1) \log\left(-H \tau_1\right) \,, \label{logt1}
    \end{align}
    where $g$ is a polynomial of $\tau_1$, with coefficients depending on $k_i$ and $p_j$. Such integrals evaluate to the following,
\begin{align}
    f_1(k_i,p_j)+\log \left( \frac{H}{k_T} \right) f_2(k_i,p_j) \,, \label{eqn:logkT}
\end{align}
    where $f_1(k_i,p_j)$ is a finite term and $f_2(k_i,p_j)$ is the result of the time integrals evaluated without any logs.

    \item \textbf{Integrals with $\mathbf{\log \left(-H \tau_2 \right)}$ :} Evaluation of $\tau_2$ integral in Eq. \eqref{eqn:2vertices} (and $\tau_3$ integral in Eq. \eqref{eqn:3vertices}) gives the following sum of exponential integral functions (Ei) and polynomials in $\tau_1$,
    \begin{align}
        &\int_{-\infty}^0 d\tau_1~ \tau_1^m e^{i a~\tau_1} \{\text{Ei}\left[i (b+c) \tau _1\right] g_1(k_i,p_j) + e^{i(b+c)~\tau_1} g_2(k_i,p_j,\tau_1) \nonumber \\
        & +e^{i(b+c)~\tau_1} \log \tau_1 ~g_3(k_i,p_j,\tau_1) \} \,,
        \label{logt2}
    \end{align}
     where $g_2$ and $g_3$ are polynomials in $\tau_1$ and $g_1$ is a function of $k_i$ and $p_j$. Upon evaluation of this integral, the first and second terms give some finite functions of $k_i$ and $p_j$. The first term also gives a log term of the form $\log \left( \frac{k_T}{p_+ + k_{12}}\right)$. There are however no additional log contributions from this piece since it converges as $p_+ \rightarrow \infty$, which implies this contribution dies at $\delta \rightarrow 0$. This was the case for e.g. in Fig~\ref{Fig:O(c_3 c_4)_1}, where the functional form of the extra log term in Eqn.~\ref{eqn:conv_q+} was such that the loop integrals were convergent.
     
     It turns out that the function $g_3$ is the result of $\tau_2$,$\tau_3$ time integrals performed without logarithms. Thus the third term in Eq. \eqref{logt2} has the same structure as Eq. \eqref{logt1}, and hence the result will have the structure of Eq. \eqref{eqn:logkT}.

     \item \textbf{Integrals with $\mathbf{\log\left(-H \tau_3\right)}$ :} Evaluation of $\tau_3$ in Eq. \eqref{eqn:3vertices} once again gives a sum of exponential integral functions and polynomials of $\tau_2$,
     \begin{align}
         & \int_{-\infty}^0 d\tau_1 \int_{-\infty}^{\tau_1}d\tau_2 ~\tau_1^m \tau_2^n \{ \text{Ei}\left[i c \tau _2\right] h_1(k_i,p_j) + e^{i c \tau_1} h_2(k_i,p_j,\tau_2) \nonumber \\
         & + e^{i c \tau_2} \log \tau_2 ~ h_3(k_i,p_j,\tau_2)\} \,,
         \label{logt3}
     \end{align}
     where $h_2$ and $h_3$ are polynomials of $\tau_2$ and $h_1$ is a function of $k_i$ and $p_j$. 

     The second term gives some finite non-log contribution when the remaining time integrals are carried out. The $\tau_2$ integral on the first term having exponential integral yields some finite answer along with more exponential integral(s), both of which give non-log finite contributions upon performing $\tau_1$ integral. The latter however also gives a $\log \left( \frac{k_T}{p_++k_{12}} \right)$, which goes to zero as $\delta \rightarrow 0$ and does not contribute any additional logs, since this term is finite at $p_+ \rightarrow \infty$.

The third term in Eq. \eqref{logt3} has the same structure as Eq. \eqref{logt2}, as the function $h_3$ is the result $\tau_3$ integral performed without the log. Hence this evaluation will once again give the same structure as Eq. \eqref{eqn:logkT}.

\end{itemize}
Hence we conclude that in all our computations of $\mathcal{O}(\delta)$ correction terms, $D_2$ is given by 
\begin{align}
   \sim \delta \log \left(\frac{H}{k_T} \right)~D_1 + ~\text{finite} \,.
\end{align}

\section{$i\epsilon$ prescription in cosmology and the in-in formalism}\label{in:in}

In this section we review the in-in formalism and the $i\epsilon$ prescription in cosmology\cite{Maldacena:2002vr,Weinberg:2005vy,Senatore:2016aui}. In Cosmology, we are interested in the late time ($\tau\rightarrow0$) correlation functions provided we start in the interacting vacuum, $\ket{\Omega(\tau_0)}$, at some early time, $\tau_0$ at the beginning of Inflation. Since the inflationary background is time-dependent, the state at late times is no longer the vacuum but a very high occupation number, semi-classical state, $\ket{\psi(0)}$. Therefore, the object of interest is,
\begin{align}
    _I\bra{\psi(0)} Q_{I}(0)  \ket{\psi(0)}_I\,,
\end{align}
where $Q_{I}(0)$ is the operator we wish to compute the expectation value for at late times. The above equation is written in the interaction picture. We do not know what $\ket{\psi(0)}$ is but we do know the initial state, therefore, we evolve backwards to the onset of inflation and get,
\begin{align}
    _I\bra{\Omega(\tau_0)} U^{\dagger}_{I}(0,\tau_0)Q_{I}(0)  U_{I}(0,\tau_0)\ket{\Omega(\tau_0)}_I\,.
\end{align}
We identify all pictures at time $\tau_0$, which means $\ket{\Omega(\tau_0)}=\ket{\Omega}$ (which is the Heisenberg picture ket) and get, 
\begin{align} \label{cosmo:inin}
    \bra{\Omega} U^{\dagger}_{I}(0,\tau_0)Q_{I}(0)  U_{I}(0,\tau_0)\ket{\Omega}\,.
\end{align}
Now, to proceed further we need to express the interacting vacuum in terms of something which we know how to calculate. In flat space QFT, we use the $i\epsilon$ prescription to project the free vacuum on the interacting vacuum in the following way, consider the object
\begin{align}
    e^{-iHT}\ket{0}=e^{-iE_0T}\ket{\Omega}\bra{\Omega}\ket{0}+\sum_{n\neq 0}e^{-iE_nT }\ket{n}\bra{n}\ket{0}\,,
\end{align}
which is obtained by expanding the free vacuum in terms of eigenstates of the interacting Hamiltonian. Now take $T\rightarrow \infty(1-i\epsilon)$, this projects $\ket{0}$ on $\ket{\Omega}$ since
\begin{align}
    \lim_{T\rightarrow \infty(1-i\epsilon)}\sum_{n\neq 0}e^{-iE_nT \ket{n}}\bra{n}\ket{0} \longrightarrow 0\,,
\end{align}
and we have,
\begin{align}
      \lim_{T\rightarrow \infty(1-i\epsilon)} e^{-iHT}\ket{0}=e^{-iE_0T}\ket{\Omega}\bra{\Omega}\ket{0}\,.
\end{align}
Notice, that the above equation seems to be telling us that the interacting vacuum at $t=0$ can be obtained from the free vacuum in the far past with slight deformation of the contour into the complex plane. In other words, it looks as if the $i\epsilon$ has turned off interactions in the far past and turned the interacting vacuum into the free vacuum. This intuition is not correct because one could have projected any other state (Any state in the free or any generic state in the interacting theory) instead of $\ket{0}$, provided that it has a non-zero overlap with the interacting vacuum. For any general state $\ket{\psi}$ (with non-zero overlap with $\ket{\Omega}$) we have,
\begin{align}
     \lim_{T\rightarrow \infty(1-i\epsilon)} e^{-iHT}\ket{\psi}=e^{-iE_0T}\ket{\Omega}\bra{\Omega}\ket{\psi}\,.
\end{align}
Therefore, it is best to only think of $i\epsilon$ as a projector for the interacting vacuum, mathematically,
\begin{align} \boxed{
     \lim_{T\rightarrow \infty(1-i\epsilon)} e^{-iHT}=e^{-iE_{0}T}\ket{\Omega}\bra{\Omega}} \,,
\end{align}
provided that the projected state has a non-zero overlap with $\ket{\Omega}$. Now, let us get back to the cosmological case, Eq. \eqref{cosmo:inin}. First, we consider the ket side (the bra side is just the adjoint), we have,
\begin{align}
    U_{I}(0,\tau_0)\ket{\Omega}\,.
\end{align}
We assume that inflation starts at a time ($\tau_0$) early enough such that the relevant modes are deep inside the horizon and because of the equivalence principle are essentially in flat space i.e. one can neglect time dependence due to the expansion. Therefore, for these early times, the interaction picture operator has the usual flat space form,
\begin{align}
    U_{I}(\tau,\tau')=e^{iH_{0}(\tau-\tau_0)}e^{-iH(\tau-\tau')}e^{-iH_{0}(\tau'-\tau_0)}\,,
\end{align}
where the interaction picture is defined at time $\tau_0$. Now, we use this object to project the free vacuum onto $\ket{\Omega}$ in the following way,
\begin{align}
      \lim_{T\rightarrow \infty(1-i\epsilon)}U_{I}(\tau_0,-T)\ket{0}&=  \lim_{T\rightarrow \infty(1-i\epsilon)}e^{-iH(\tau_0+T)}e^{-iH_{0}(-T-\tau_0)}\ket{0}  \nonumber \\
     & = \lim_{T\rightarrow \infty(1-i\epsilon)}e^{-iH(\tau_0+T)}\ket{0} \nonumber \\
     & =  \lim_{T\rightarrow \infty(1-i\epsilon)}e^{-iE_0(T+\tau_0)}\ket{\Omega}\bra{\Omega}\ket{0}+\sum_{n\neq 0}e^{-iE_n(T+\tau_0)} \ket{n}\bra{n}\ket{0} \,,
\end{align}
where, we used the fact that $H_0\ket{0}=0$. Again like in the previous case, the summation goes to zero and we get,
\begin{align}\label{ket}
        \lim_{T\rightarrow \infty(1-i\epsilon)}U_{I}(\tau_0,-T)\ket{0}=e^{-iE_0(T+\tau_0)}\ket{\Omega}\bra{\Omega}\ket{0}\,.
\end{align}
The bra side is the adjoint of this,
\begin{align}\label{bra}
       \lim_{T\rightarrow \infty(1+i\epsilon)}\bra{0}U^{\dagger}_{I}(\tau_0,-T)=e^{iE_0(T+\tau_0)}\bra{\Omega}\bra{0}\ket{\Omega}\,.
\end{align}
Put Eq. \eqref{ket} and Eq. \eqref{bra} in Eq. \eqref{cosmo:inin},
\begin{align}
    \bra{0}U^{\dagger}_{I}(\tau_0,-\infty^{+}) U^{\dagger}_{I}(0,\tau_0)Q_{I}(0)  U_{I}(0,\tau_0)U_{I}(\tau_0,-\infty^{-})\ket{0} \,,
\end{align}
which gives the final relation,
\begin{align} \boxed{
     \bra{0}U^{\dagger}_{I}(0,-\infty^{+})Q(0)U_{I}(0,-\infty^{-})\ket{0} }\,,
\end{align}
where, $\infty^{\pm}=\infty(1\pm i\epsilon)$.

\bibliography{reference.bib}

\end{document}